\documentclass[aps,prl,preprint,superscriptaddress]{revtex4-2}

\usepackage[dvips] {graphics}
\usepackage {epsfig}
\usepackage[mathscr]{eucal}
\usepackage{amsfonts}
\usepackage{amsmath}
\usepackage{pgf,xcolor}
\usepackage{dsfont}

\usepackage[american]{babel}
\usepackage{bm}

\newcommand{\bk}{\mathbf{k}}
\newcommand{\eps}{\varepsilon}
\newcommand{\ho}{\hbar \, \omega}
\newcommand{\sv}{\, ,}

\newcommand{\BE}{\begin{equation}}
	\newcommand{\EE}{\end{equation}}

\newcommand{\R}{\mathbb{R}}

\begin{document}


\title{The Pauli principle and the Monte Carlo Method for charge transport in graphene}


\author{Marco Coco}
\email[]{m.coco@univpm.it}
\affiliation{Dipartimento di Ingegneria Industriale e Scienze Matematiche, Universit\`{a} Politecnica delle Marche, via Brecce Bianche, 12, 
	60131 Ancona, Italy }

\author{Paolo Bordone}
\email[]{paolo.bordone@unimore.it}
\affiliation{Dipartimento di Fisica, Informatica e Matematica, Universit\`{a} di Modena e Reggio Emilia, via Campi 213/A, 41125 Modena, Italy }

\author{Lucio Demeio}
\email[]{l.demeio@univpm.it}
\affiliation{Dipartimento di Ingegneria Industriale e Scienze Matematiche, Universit\`{a} Politecnica delle Marche, via Brecce Bianche 12, 60131 Ancona, Italy }

\author{Vittorio Romano}
\email[]{romano@dmi.unict.it}
\affiliation{Dipartimento di Matematica e Informatica, Universit\`{a} degli Studi di Catania, viale Andrea Doria 6, 95125, Catania, Italy}


\date{\today}

\begin{abstract}
The attempt to include the Pauli principle in the Monte Carlo method by acting also on the free flight step and not only at the end of each collision is investigated. The charge transport in suspended monolayer graphene is considered as test case. The results are compared with those obtained in the standard Ensemble Monte Carlo technique and in the new Direct Simulation Monte Carlo algorithm which is able to correctly handle with Pauli's principle. The physical aspects of the investigated approach are analyzed as well.
\end{abstract}
%

\maketitle

\section{Introduction}
\label{Intr}
The Monte Carlo approach is nowadays of widespread use in the simulations of charge transport in semiconductor devices. Most of the methods employed in this field are based on  the Ensemble Monte Carlo (EMC) procedure of Lebwohl and Price \cite{LP}, and on the approach developed by Bosi, Jacoboni and Reggiani \cite{BJ}-\cite{JacReg}. When the simulations involve degenerate materials, the inclusion of the Pauli principle becomes essential; to this aim, Lugli and Ferry \cite{LuFe} improved the EMC method and included the Pauli principle by means of a rejection technique at the end of each scattering process. Unfortunately, with this procedure the charge distribution can exceed the maximum value of one, leading to unphysical results. Some attempts were made in the literature at overcoming this difficulty: in \cite{BoTho,BoAda} ad hoc scattering out terms to force the distribution function to be smaller than one were introduced; in \cite{FisLax,ZeBuEsSh} some approximations of the distribution of the final states were used. These efforts improved the simulation results but did not lead to a correct reconstruction of the distribution function, that is of fundamental importance, since it enters the collisional terms of the Boltzmann equation and influences the determination of the scattering probabilities. 

An important issue, which we address in this paper, arises when degenerate materials are considered. In this case, there is an ongoing debate in the literature on whether the Pauli principle should be applied to the free flight step or not. This is an important theoretical question, also for computational purposes. An alternative procedure was proposed in \cite{Tady} where the rejection technique was adopted not only at the end of each scattering event but also at the end of each free flight. This method, although having been cited several times in the literature (see for example \cite{BoTho,BoAda,ZeBuEsSh,1,3}), to the best of our knowledge it has never been used. In the literature an analysis of this approach is missing (it has received only a brief comment in \cite{BoTho}); it is the aim of this work to fill this gap and to present a coherent discussion on the inclusion of the Pauli principle in a Monte Carlo procedure. We do so by comparing the numerical results for a suspended monolayer graphene obtained with the Monte Carlo Method presented in \cite{Tady} with those obtained by using the standard EMC in \cite{LuFe} and the new Direct Simulation Monte Carlo (DSMC) in \cite{RMC}, that are by now well established in the semiconductors field and cross-validated with deterministic solutions, for example those based on the discontinuous Galerkin method (\cite{RMC}, \cite{CMR_Ric}-\cite{MNR}) or on WENO schemes \cite{LiMo}. 

The new DSMC scheme in \cite{RMC} is able to correctly include the Pauli principle; the streaming term of the Boltzmann equation is treated deterministically by means of a splitting procedure, resulting in a rigid translation of the distribution function as a whole. The scattering events are then simulated and the rejection technique is applied at the end of each collision. 

The results obtained by using the EMC and the new DSMC are already studied in \cite{RMC} and allows us to quantify the correctness of the method proposed in \cite{Tady}. Besides, graphene is a material whose peculiar energy bands make the degeneracy effects relevant, so representing a useful choice as test case.

The paper is organized as follows. In section \ref{model} the semiclassical mathematical model for spatially homogeneous graphene is presented and the simulation procedures are introduced; in sections \ref{results} we present and discuss the results of our simulations; section \ref{conclusions} contains our conclusions. \\

\section{The mathematical model and the Monte Carlo techniques}
\label{model}
The carrier population in graphene is made of four components: the electrons of the valence and of the conduction band, which can occupy the states around either of the Dirac points, $K$ and $K'$, of each band. In our simulations, we consider a homogeneous graphene strip of infinite extension in the direction transversal to the applied electric field and only the electrons of the conduction band belonging to the valley around the $K$ point, by considering all valleys as equivalent. We recall that the graphene Brillouin zone $\cal{B}$ has hexagonal shape, and we shall choose the reference frame of the $\bf k$-space with the origin to coincide with the $K$ point. The electric field is directed along the $x$-axes.

With good approximation \cite{CaNe}, the dispersion relation for the band energy $\eps$ around the equivalent Dirac points is given by
\begin{equation}\eps =   \hbar \, v_F \left| \bk -\bk_{\ell} \right|, \label{electron_dispersion}
\end{equation} 
where $v_F$ is the Fermi velocity and $\bk_{\ell}$ is the position of the Dirac point $\ell$. We will use Eq. (\ref{electron_dispersion}) as dispersion relation because for the electric field strengths usually considered in the applications the charge transport involves almost exclusively the electrons around the $K$ and $K'$ points.

Under these conditions and by using the semiclassical approximation, the Boltzmann equation for the charge carriers is

\begin{eqnarray}
	\dfrac{\partial f(t,\bk)}{\partial t} - \dfrac{e}{\hbar} \, E \, \frac{\partial f(t,\bk)}{\partial k_x} = \left. \dfrac{df}{d t}(t,\bk) \right|_{e-ph} \sv \label{bulk1}
\end{eqnarray}
where $f(t,\bk)$ is the electron distribution function of the charge carriers at time $t$, ${\bk}=\left( k_x,k_y\right)$ is the wave-vector and $\nabla_{\bk}$ is the gradient with respect to $\bk$. The r.h.s. of Eq. \ref{bulk1} is the collision operator which describes the interactions of the carriers with the phonons. 

The appropriate initial condition for Eq. \ref{bulk1} in the degenerate case is the Fermi-Dirac distribution

\begin{eqnarray}
	f(0,\mathbf{k})=f_{FD}(\mathbf{k})\equiv\frac{1}{1+\exp\left( \frac{\varepsilon(\mathbf{k})-\varepsilon_F}{k_B T}\right) }\,,
\end{eqnarray}
where $\varepsilon_F$ is the Fermi level and $T$ the room temperature, related to the charge distribution by means of

\begin{eqnarray}
	\rho(0)=\frac{2}{(2 \pi)^2}\int f(0,\mathbf{k}) d^2 \mathbf{k} \label{cd}
\end{eqnarray}
where only the spin degeneracy is considered. The Fermi level will be chosen high enough in order to produce a strong degeneracy, according to the situation we want to investigate. This is equivalent to introduce a high n-doping in a traditional semiconductor.

The electron mean energy and velocity are defined as

\begin{eqnarray}
	{\cal{E}}(t)=\frac{1}{\rho(t)}\frac{2}{(2 \pi)^2}\int \varepsilon(t,\mathbf{k})f(t,\mathbf{k}) d^2 \mathbf{k}
\end{eqnarray} 

\begin{eqnarray}
	\mathbf{V}(t)=\frac{1}{\rho(t)}\frac{2}{(2 \pi)^2}\int \mathbf{v}(t,\mathbf{k})f(t,\mathbf{k}) d^2 \mathbf{k}\,,
\end{eqnarray} 
where $\varepsilon(t,\mathbf{k})$ and $\mathbf{v}(t,\mathbf{k})$ are the particle energy and velocity, respectively, and $\rho(t)$ the time dependent electron density

\begin{eqnarray}
	\rho(t)=\frac{2}{(2 \pi)^2}\int f(t,\mathbf{k}) d^2 \mathbf{k} \label{cd2}
\end{eqnarray}

The collision operator represents the interactions of the electrons with acoustic, optical and $K$ phonons. 
Acoustic phonon scattering is intra-valley and intra-band and can be longitudinal ({\em LA}) or transversal ({\em TA}). Optical phonon scattering is 
intra-valley and can be longitudinal ({\em LO}) and transversal ({\em TO}); it can be 
intra-band, leaving the electrons in the same band, or inter-band, pushing the electrons from 
the initial band toward another one. Scattering with  $K$-phonons pushes electrons from 
a valley to a nearby one (inter-valley scattering).
For the optical and $K$ phonons we will assume the Einstein approximation, $\hbar \omega_{A}=\mbox{const}$, $A = LO, TO, K$, with $\omega_A$ the $A$th phonon frequency. The $K$ phonons are not an actual physical phonon branch, their name being due to the fact that their wave-vectors are close to the $K$ or $K'$ 
point \cite{CaNe}. They belong to the optical branches and induce intervalley scatterings. This justifies the use of the Einstein approximation for them. For the in plane  acoustic phonons the Debye approximation will be adopted, $\hbar \omega_{A} = \hbar v_A \left|\bf q \right| $, $A = LA, TA$, with $\bf q \in \cal{B}$ the phonon wave-vector, and in this case for the analytical calculations the Brillouin zone can be consistently extended to $\R^2$. The out of plane $Z$ phonons are not considered because they do not interact with the electrons; they are important when phonon transport and thermal effects are taken into account (\cite{CR}-\cite{CR_19}). 

The general form of the collision term can be written as (\cite{LiMo,CaNe})
\begin{eqnarray}
	\left. \dfrac{d f}{d t}(t,\bk) \right|_{e-ph} &=& \int_{\cal{B}} S(\bk', \bk) \, f(t,\bk') \left( 1 - f(t,\bk) \right) d \bk' \\ \nonumber
	&&- \int_{\cal{B}} S(\bk, \bk') \, f(t,\bk) \left( 1 - f(t,\bk') \right) d \bk' \,,
\end{eqnarray}
where $S(\bk', \bk) $ is the total transition rate and is given by the sum over the types of 
scatterings described above
\begin{widetext}
\begin{eqnarray}
	S(\bk', \bk)\! =\!
	\sum_{A} \left| G^{(A)}(\bk', \bk) \right|^{2} \!
	\left[  \left( g^{-}_{A} + 1 \right)
	\delta\! \left( \eps(\bk) - \eps(\bk') + \hbar \, \omega_A
	\right)  \right. 
	 \left. + g^{+}_{A} \,
	\delta \left( \eps(\bk) - \eps(\bk') - \hbar \, \omega_A
	\right) \right] \,,  \label{Scatt}
\end{eqnarray}
\end{widetext}
where the index $A$ runs over the phonon modes. The $\left| G^{(A)}(\bk', \bk)\right|^{2} $'s are the electron-phonon coupling matrix elements, which describe the interaction mechanism of an electron with an $A$th phonon, from the state of wave-vector $\bk'$ to the state of wave-vector $\bk$. The symbol $\delta$ denotes the Dirac delta function and $g_A ({\bf q})$ is the phonon distribution for the $A$-type phonons. In (\ref{Scatt}), $g_A^{\pm} = g_A \left( {\bf q}^{\pm} \right)$, where ${\bf q}^{\pm}  = \pm \left(\bk' - \bk \right)$, stemming from the momentum conservation. 

For both longitudinal and transversal  acoustic phonons,  we consider the elastic approximation according to which the transition rate is given by \cite{Sarma}
\begin{widetext}
\begin{equation}
	S_{A}(\bk', \bk) =	\dfrac{1}{(2 \, \pi)^{2}} \, 
	\dfrac{\pi \, D_{ac}^{2} \, k_{B} \, T}{2 \hbar \, \sigma_m \, v_{s}^{2}}
	\left( 1 + \cos \vartheta_{\bk \sv \bk'} \right) \delta \left(\eps (\bk') - \eps (\bk) \right) , \quad A = LA, TA\,,
	\label{transport_acoustic}
\end{equation}
\end{widetext}
where $D_{ac}$ is the acoustic phonon coupling constant (also called acoustic phonon deformation potential), $\sigma_m$ is the graphene areal density, $v_A$ the sound speed of the Ath acoustical phonon branch in graphene and $\vartheta_{\bk \sv \bk'}$ is the convex angle between $\bk$ and ${\bk'}$. 

The electron-phonon coupling matrix elements  of the longitudinal optical ($LO$), transversal optical ($TO$) and ${K}$ phonons are \cite{LiMo}
\begin{widetext}
\begin{eqnarray}
	\left| G^{(LO)}(\bk', \bk) \right|^{2} & = & 
	\dfrac{1}{(2 \, \pi)^{2}} \, \dfrac{\pi \, D_{O}^{2}}{\sigma_m \, \omega_{O}}
	\left( 1 - \cos ( \vartheta_{\bk \sv \bk' - \bk} + \vartheta_{\bk' \sv \bk' - \bk} ) \right)\,,
	\\
	\left| G^{(TO)}(\bk', \bk) \right|^{2} & = & 
	\dfrac{1}{(2 \, \pi)^{2}} \, \dfrac{\pi \, D_{O}^{2}}{\sigma_m \, \omega_{O}}
	\left( 1 + \cos ( \vartheta_{\bk \sv \bk' - \bk} + \vartheta_{\bk' \sv \bk' - \bk} ) \right)\,,
	\\
	\left| G^{(K)}(\bk', \bk) \right|^{2} & = & 
	\dfrac{1}{(2 \, \pi)^{2}} \, \dfrac{2 \pi \, D_{K}^{2}}{\sigma_m \, \omega_{K}}
	\left( 1 - \cos \vartheta_{\bk \sv \bk'} \right) \,,
\end{eqnarray}
\end{widetext}
where $D_{O}$ is the optical phonon coupling constant, $\omega_{O}$ the optical phonon frequency,
$D_{K}$ is the $K$ phonon coupling constant and $\omega_{K}$ the $K$ phonon frequency.
The angles $\vartheta_{\bk \sv \bk' - \bk}$ and $\vartheta_{\bk' \sv \bk' - \bk}$ denote 
the convex angles between $\bk$ and $\bk' - \bk$  and between $\bk'$ and  $\bk' - \bk$, 
respectively. 

\begin{figure}
	\centering
	\includegraphics[width=0.9\columnwidth]{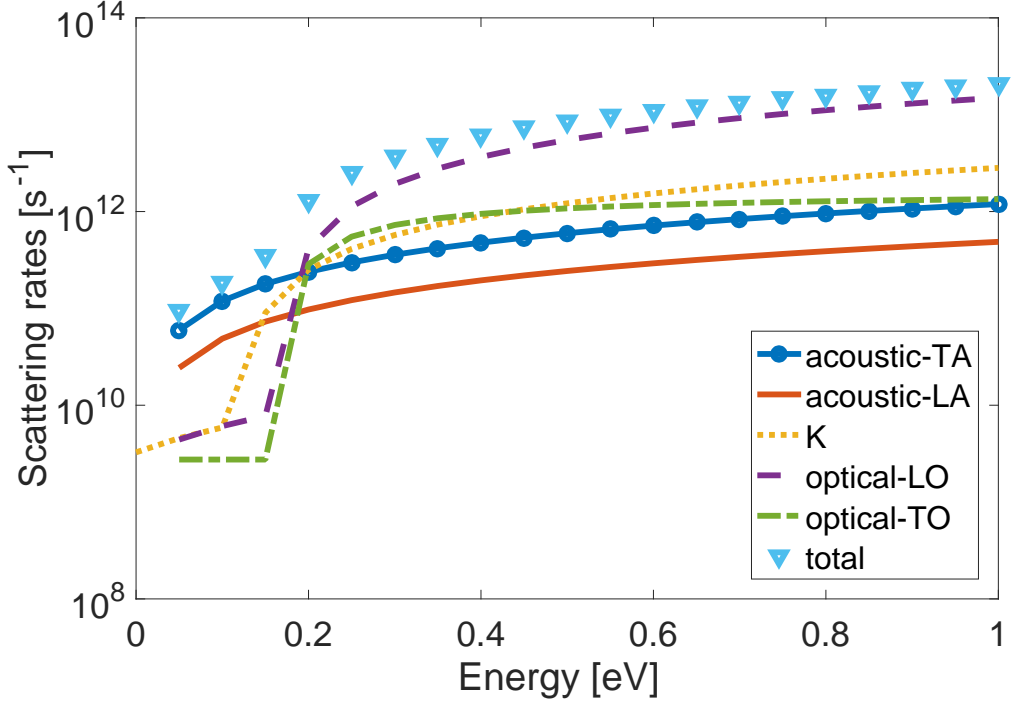}
	\caption{Scattering rates evaluated with the Bose-Einstein equilibrium distributions for phonons at the temperature of 300 K.	\label{fig:scattering_rate}}
\end{figure}

In the Monte Carlo procedure, the scattering rate of each type of scattering plays a fundamental role; the scattering rate $\Gamma_A$ of the Ath type of scattering is defined as
\begin{equation}
	\Gamma_A (\bk) =  \int_{\cal{B}} S_A (\bk, \bk') \: d \bk'. 
\end{equation}
For the sake of completeness, we provide here the scattering rates of all types of scatterings used in this paper. For the acoustic phonon scattering we get
\begin{eqnarray}
	\Gamma_{ac}  (\eps) = \dfrac{ D_{ac}^{2} \, k_{B} \, T}{4 \hbar^3  \,  v_F^2 \, \sigma_m \, v_{p}^{2}} \,\, \eps \label{rate_ac}\,;
\end{eqnarray}
for the longitudinal and transversal optical phonons
\begin{widetext}
\begin{eqnarray}
	\Gamma_{LO,TO}  (\eps) &=& \dfrac{D_{O}^{2}}{ 4 \pi \sigma_m \, \omega_{O} \hbar^2 \, v_F^2} \left[ \left(\eps -   \hbar \, \omega_{O} \right) \left( g_{LO,TO} + 1 \right) H (\eps  -   \hbar \, \omega_{O})\left(2 \pi \mp  \Lambda^{-}(\eps)\right) \nonumber \right.\\ 
	&+&   \left(\eps + \hbar \, \omega_{O} \right) 
	\,g_{LO,TO} \left( 2 \pi \mp \Lambda^{+}(\eps) \right)  \left. \right]\,,  \label{rate_op}
\end{eqnarray}
\end{widetext}
where the upper and the lower signs refer to the $LO$ and $TO$ phonons, respectively.
$H$ is the Heaviside function and
\begin{widetext}
\begin{eqnarray}
	\Lambda^{\pm}(\eps) &=& \int_0^{2 \pi}  \frac{(2 \eps^2 + \hbar^2 \omega_{O}^2 \pm 2 \hbar \omega_{O} \eps) \, \cos \theta''
		-2 \eps (\eps \pm \ho_{O})}{2 \eps^2 +  \hbar^2 \omega_{O}^2 \pm 2 \hbar \omega_{O} \eps - 2 \eps 
		(\eps \pm \ho_{O}) \cos \, \theta'' } \, d \, \theta'' \nonumber \\ 
	&=& -\frac{\pi}{\eps (\eps \pm \hbar \omega_{O})} \left( 2 \eps^2+\hbar^2 \omega_{O}^2 \pm 2 \hbar \omega_{O} \eps - \hbar \omega_{O} |2 \eps \pm \hbar \omega_{O} | \right)\,.
\end{eqnarray}
\end{widetext}
Similarly, the expression of the scattering rate for the $K$ phonon scattering reads
\begin{widetext}
\begin{eqnarray}
	\Gamma_{K}  (\eps) = \dfrac{D_{K}^{2}}{ \sigma_m \, \omega_{K} \hbar^2 \, v_F^2} \left[\left(\eps -   \hbar \, \omega_{K} \right) \left(g_{K} + 1 \right) H (\eps -   \hbar \, \omega_{K}) +        \left(\eps + \hbar \, \omega_{K} \right) \, g_{K}\right]. \label{rate_K}
\end{eqnarray}
\end{widetext}
These scattering rates are shown in Fig.~\ref{fig:scattering_rate} as functions of the energy.

In this work, we compare the results for charge transport in graphene obtained with three different Monte Carlo procedures: the standard Ensemble MC procedure, a new Ensemble MC procedure and the Free Flight based one. 

In general, in a Monte Carlo simulation, the motion of each particle is given by the solution of the semi-classical equation of motion $\hbar\dot{\bf k}=-e{\bf E}$, followed by a collisional event after a time $\Delta t$. The total scattering rate is given by 
\begin{eqnarray}
	\tilde{\Gamma}_{tot}=\Gamma_{LA}+\Gamma_{TA}+\Gamma_{LO}+\Gamma_{TO}+\Gamma_{K} \,,
\end{eqnarray}
and it varies with the energy. A scattering rate $\Gamma_{ss}$ due to fictitious collision events, called self-scatterings and which do not change the particle state, is also introduced and a new constant total scattering rate is defined as $\Gamma_{tot}=\tilde{\Gamma}_{tot}+\Gamma_{ss}$. $\Delta t$ is then calculated as the ratio (see for example \cite{JL})
\begin{eqnarray}
	\Delta t= -\frac{\ln \eta}{\Gamma_{tot}}\,, \label{deltat}
\end{eqnarray}
where $\eta$ is a random number uniformly distributed in $[0,1]$. In general, the value of $\Gamma_{tot}$ is determined as $\alpha \Gamma_{max}$, being $\Gamma_{max}=\max\left(\Gamma_{LA}+\Gamma_{TA}+\Gamma_{LO}+\Gamma_{TO}+\Gamma_{K}\right)$, with $\alpha >1$ a tuning parameter. Since the previous scattering rates can differ even by two order of magnitude, using the same $\Gamma_{tot}$ for each time step leads to a very large number of self-scatterings, making the computational cost considerably great. Therefore, in our simulations, we use a variable $\Gamma_{tot}$ depending on the state of the particle at the current time $t$:
\begin{widetext}
\begin{eqnarray}
	\Gamma_{tot}=\alpha\left(\,\Gamma_{LA}(\varepsilon(t))+\Gamma_{TA}(\varepsilon(t))+\Gamma_{LO}(\varepsilon(t))+\Gamma_{TO}(\varepsilon(t))+\Gamma_{K}(\varepsilon(t)) \, \right)\,,
\end{eqnarray} 
\end{widetext}
with $\alpha=1.1$. 

In our simulations, the following three Monte Carlo procedures will be used:

	{{\it (i) the Standard Ensemble Monte Carlo (SEMC)}} \cite{LuFe}.\\
	After each collision, the new wave-vector $\mathbf{k'}$ is determined and if the final state is available the initial state $\mathbf{k}$ is updated. The availability of the final states has to respect the Pauli principle and this is checked by means of a rejection procedure: a random number $\zeta$, uniformly distributed in $[0,1]$, is generated and the final state $\mathbf{k'}$ is available if the condition $\zeta<1-f(\mathbf{k'})$ holds. This scheme is repeated for each particle. 
	
	{{\it (ii) the New Ensemble Monte Carlo (NEMC)}} \cite{RMC}.\\
	It consists of two main steps. First, the distribution function is translated as a whole according to the semi-classical equation of motion, and all the particles experience the same free flight; then, for each particle a sequence of collisional events is simulated. The final state after each scattering mechanism is checked by using the rejection technique described above. 
	
	{{\it (iii) the Free Flight based Monte Carlo (FFMC)}} \cite{Tady}.\\
	In this method, the rejection technique to check the availability of the final states is used not only at the end of each collision but also at the end of each free flight; if the state reached after the free flight governed by the semi-classical equation of motion is not accepted, the particle goes back and nothing happens.

\section{Simulation results}
\label{results}

In this section, the results of the three previous procedures to include the Pauli principle in the Monte Carlo method are shown, compared and discussed.

In the simulations, $n_P=10^4$ (super)-particles are used, the time step is set equal to $\Delta t=2.5$ fs; the wave-vector space is discretized by means of a uniform square grid $[-k_{x max}, k_{x max}]\times[-k_{y max}, k_{y max}]$, with $k_{x max}=k_{y max}=24 \, \mbox{nm}^{-1}$,  and $642 \times 642$ cells.  The physical parameters proposed in \cite{Li2010, Bor} and reported in Table \ref{table1} are adopted.
 \begin{table}
	\caption{Physical parameters for the scattering rates.\label{table1}}
	\centering
	\begin{ruledtabular}
		\begin{tabular}{c|c}
			$\sigma_m$ & $7.6 \times 10^{-8}$ g/cm$^2$\\
			$v_F$         & $10^6$ m/s\\ 
			$v_{LA}$       & $ 2.13 \times 10^4$ m/s\\
			$v_{TA}$       & $ 1.36 \times 10^4$ m/s\\
			$D_{ac}$    & $6.8$ eV \\ 
			$ \hbar \, \omega_{LO}$ & $164.6$ meV\\
			$ \hbar \, \omega_{TO}$ & $164.6$ meV\\
			$D_{O}$    & $10^9$ eV/cm\\ 
			$ \hbar \,\omega_{K}$ & $124$ meV \\
			$D_{K}$ & $3.5 \times 10^8$ eV/cm\\
		\end{tabular}
	\end{ruledtabular}
\end{table}

\begin{figure}[h!]
	\centering
	\fbox {a)		\includegraphics[width=0.8\columnwidth]{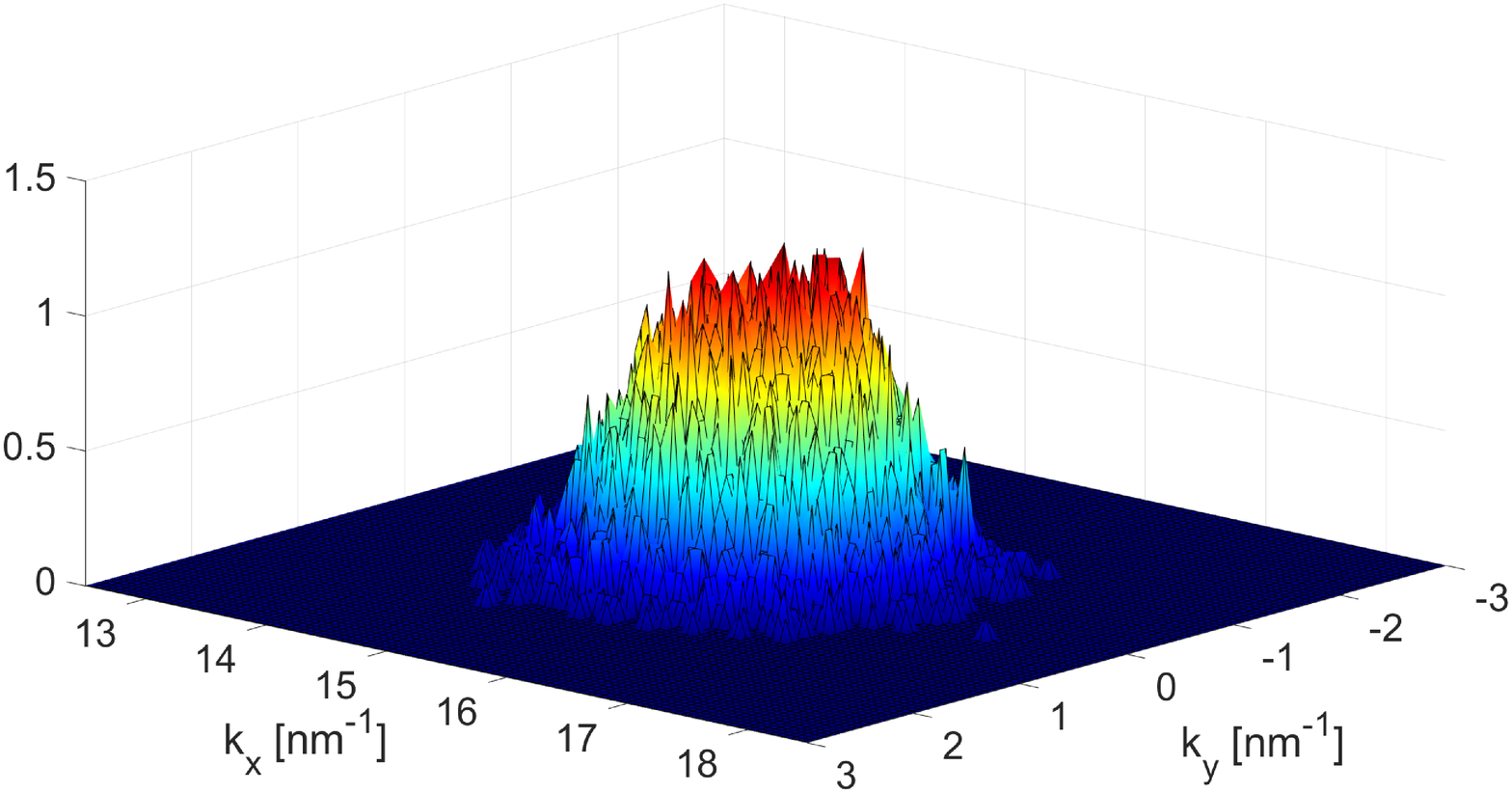}}\\
	\fbox {b)		\includegraphics[width=0.8\columnwidth]{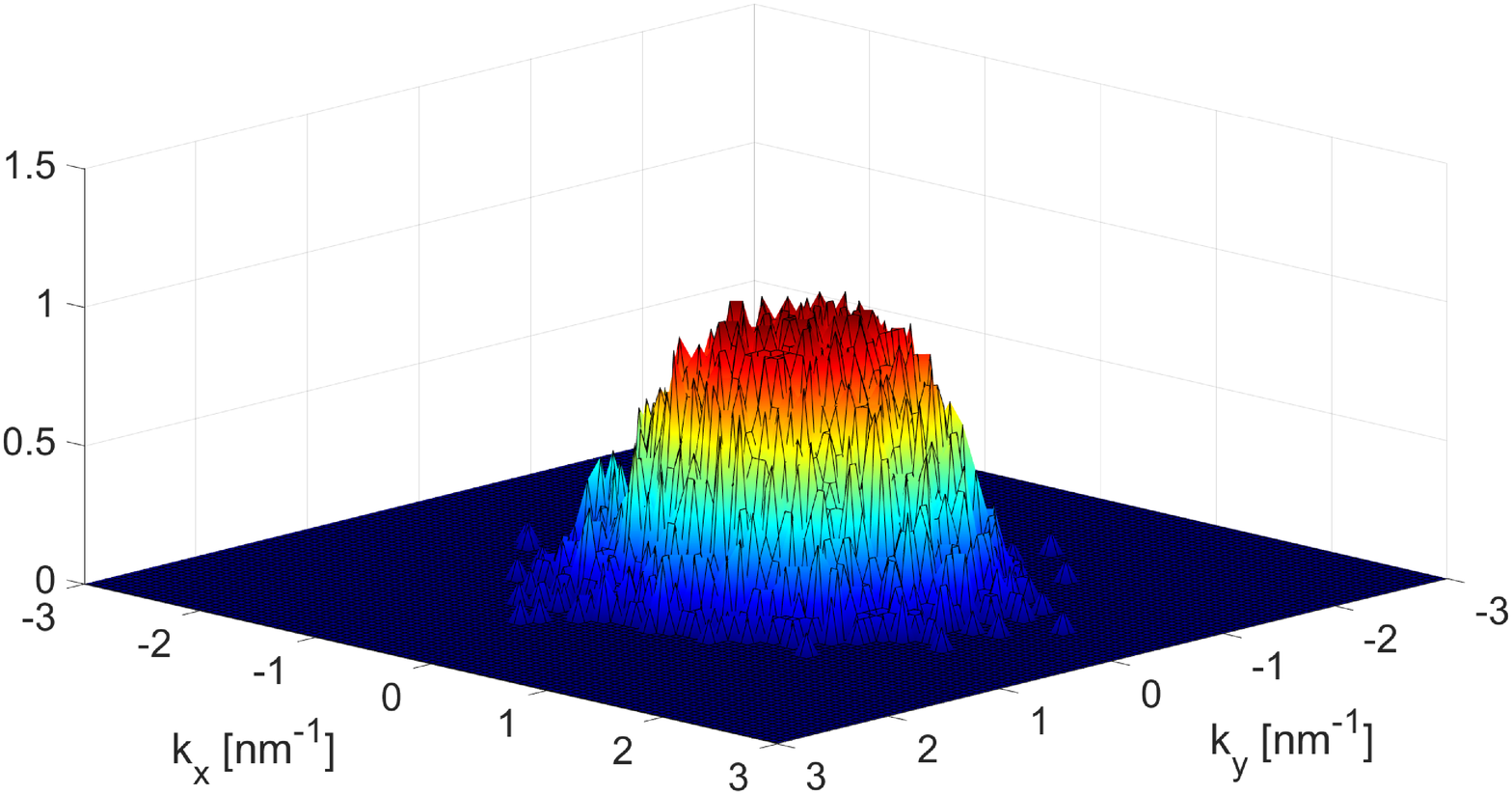}}\\
	\caption{Distribution functions after $5 \, \mbox{ps}$ for SEMC, a), and NEMC, b), with $\varepsilon_F=0.6$ eV and $E=20$ kV/cm.	\label{fig:EMC_DSMC_distr}}
\end{figure}

\begin{figure}[h!]
	\centering
	\fbox {a)		\includegraphics[width=0.41\columnwidth]{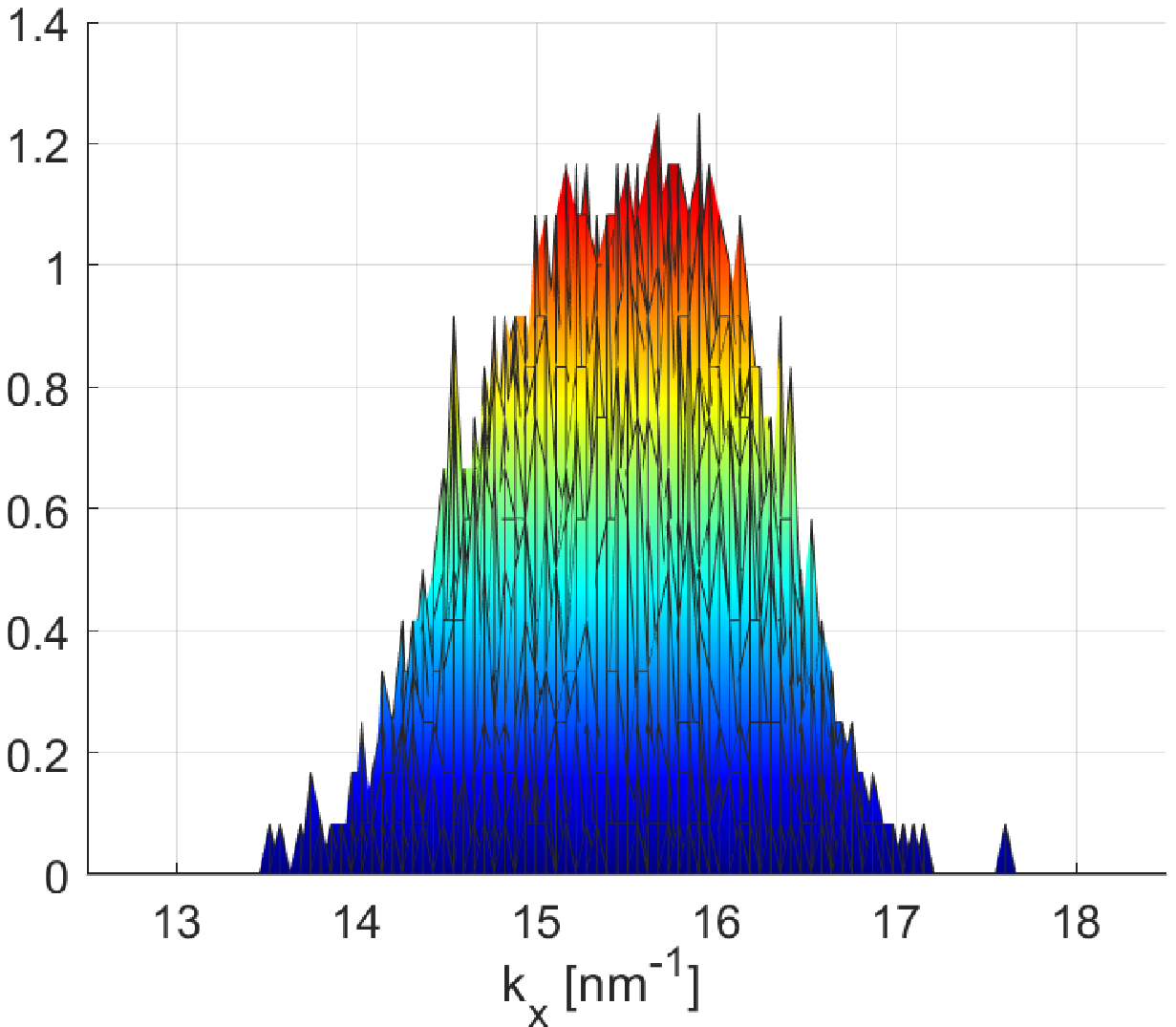}}
	\fbox {b)		\includegraphics[width=0.41\columnwidth]{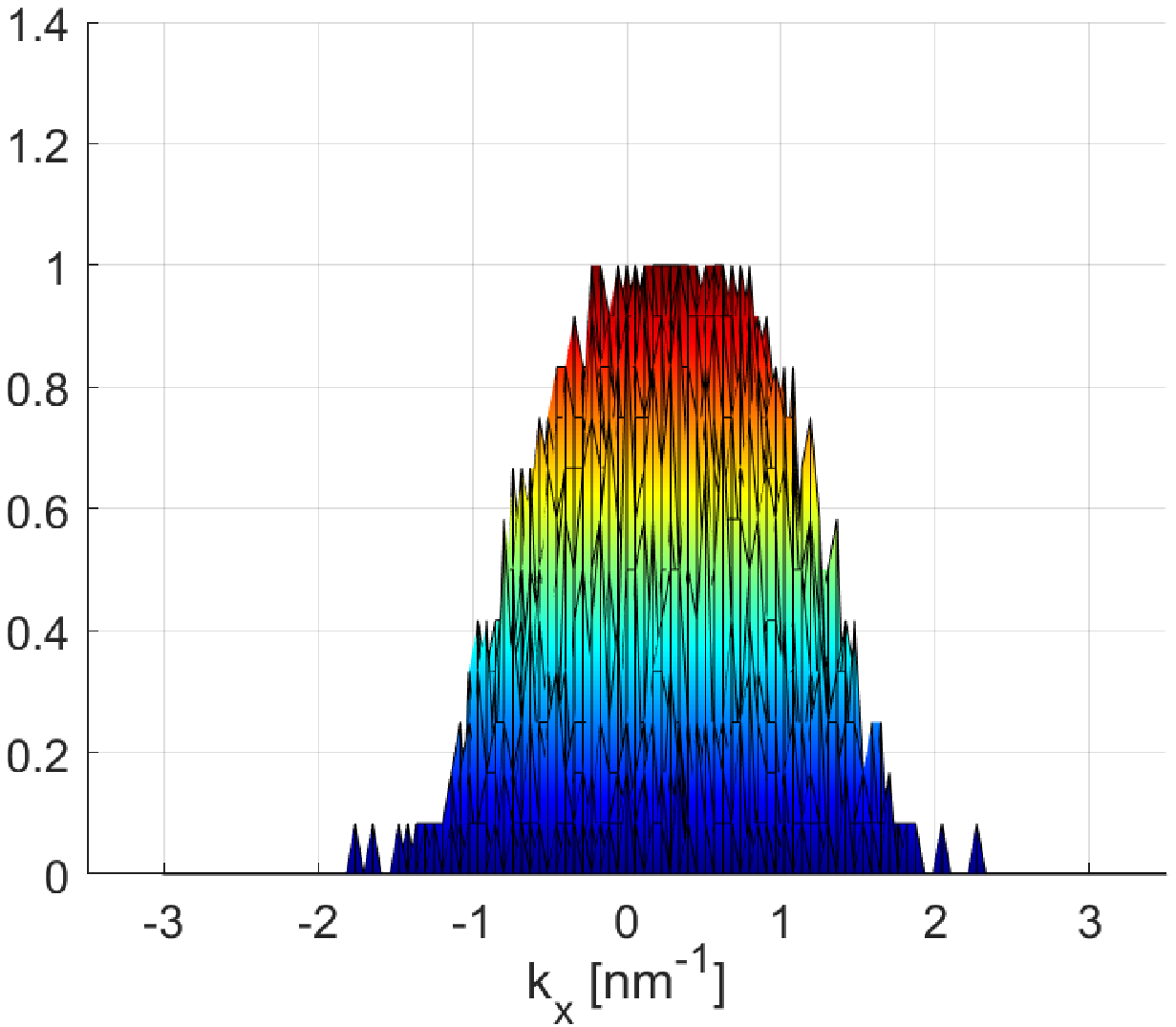}}\\
	\caption{View of the distribution functions along the x-axis after $5 \, \mbox{ps}$ for SEMC, a), and NEMC, b), with $\varepsilon_F=0.6$ eV and $E=20$ kV/cm.	\label{fig:EMC_DSMC_x_view}}
\end{figure}

\begin{figure}
	\centering
	\fbox {a)		\includegraphics[width=0.41\columnwidth]{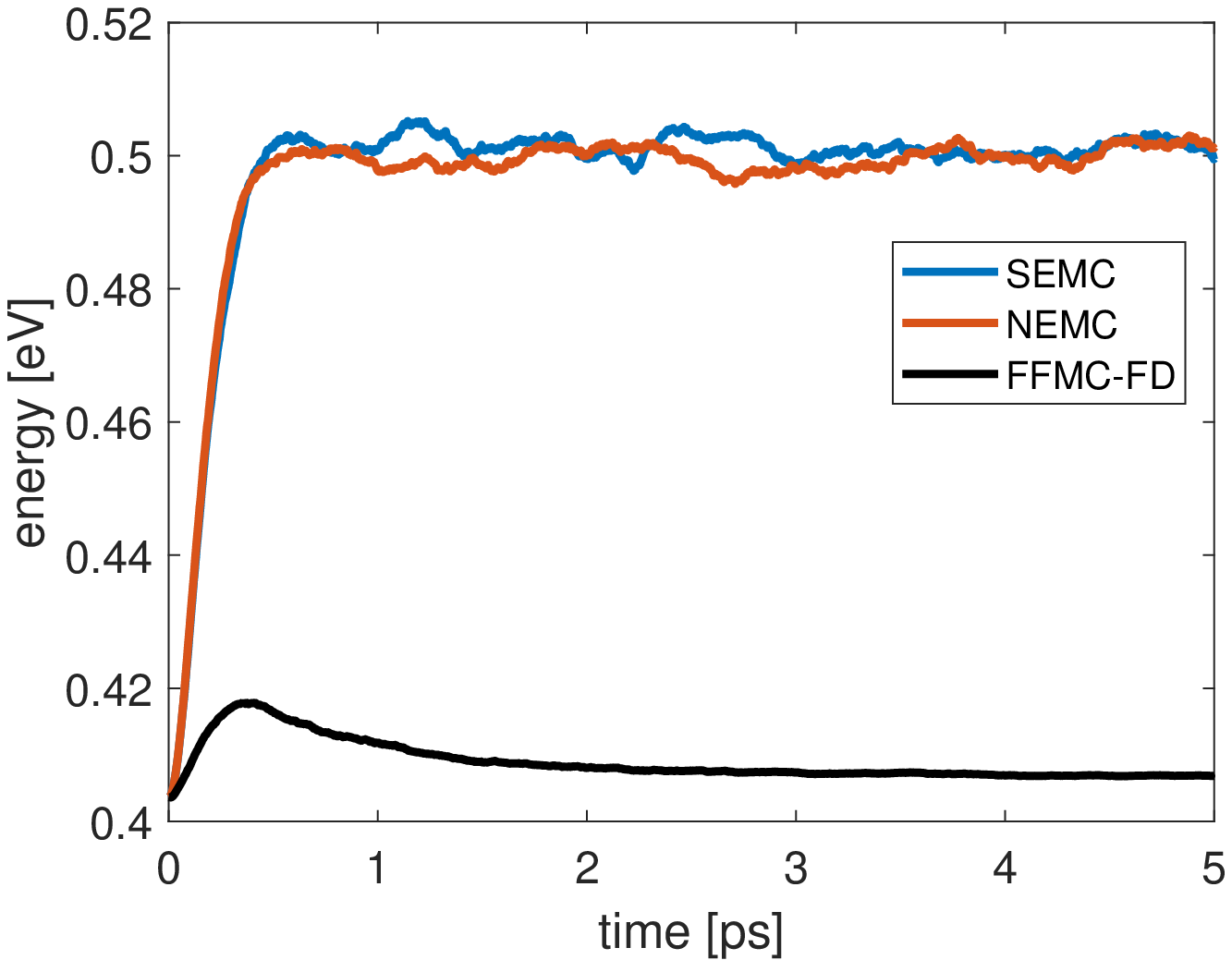}}
	\fbox {b)		\includegraphics[width=0.41\columnwidth]{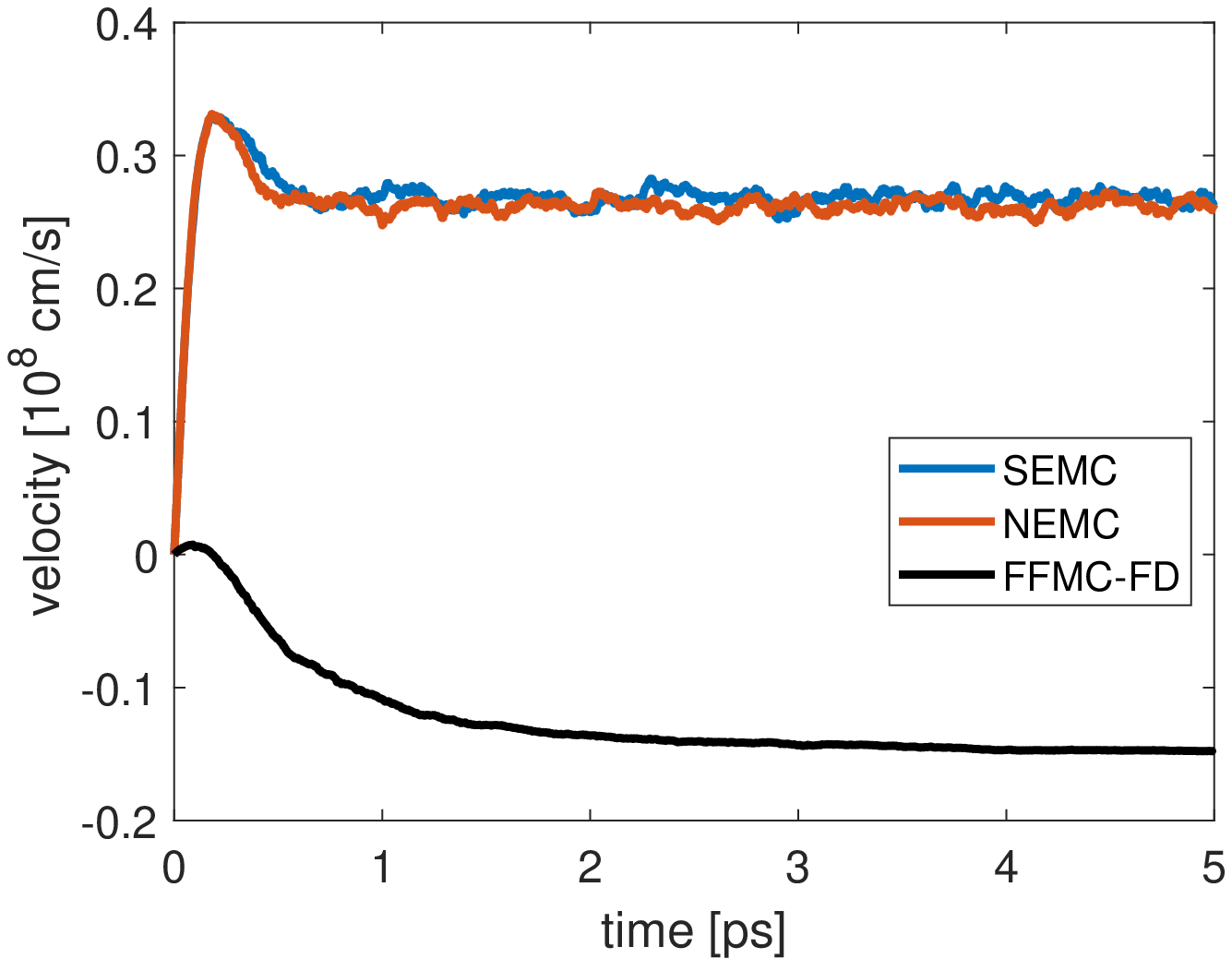}}\\
	\caption{Mean energy a) and velocity b) for SEMC, NEMC, FFMC procedures, with $\varepsilon_F=0.6$ eV and $E=20$ kV/cm.	\label{mean_FFMC}}
\end{figure}

Fig. \ref{fig:EMC_DSMC_distr} shows the particle distribution for the SEMC (a) and the NEMC (b), techniques. The NEMC procedure is able to properly take into account the Pauli exclusion principle and the distribution function is correctly reconstructed and bounded between zero and one (Fig. \ref{fig:EMC_DSMC_distr} b), unlike the SEMC method (Fig. \ref{fig:EMC_DSMC_distr} a), with which the distribution takes values larger than one. The section of the distributions along the $x$ direction in Fig. \ref{fig:EMC_DSMC_x_view} highlights the difference between the results in the two cases. In Fig. \ref{mean_FFMC} the mean energy and velocity for both schemes are reported; the results of the NEMC and SEMC methods are in good agreement (for further details see \cite{RMC}, where a cross validation with the deterministic results obtained by means of the Discontinuous Galerkin method is analyzed as well). The x-axes of Figs. \ref{fig:EMC_DSMC_distr}-\ref{fig:EMC_DSMC_x_view} show how the grid is rigidly moved together with the whole distribution function, consistently with the semi-lagrangian approach of the NEMC.

In Fig. \ref{mean_FFMC} the average energy and velocity obtained with the FFMC procedure are shown and compared with those obtained with SEMC and NEMC. The energy remains almost constant and the velocity, after an initial negligible positive peak, reaches a negative value equal to $-0.14\times10^8 \,\mbox{cm/s}$. The behavior given by the FFMC approach is certainly unphysical because it is opposite to the one induced by the electric field. In \cite{Tady}, this result is explained as a consequence of the initial Fermi-Dirac condition which leads to have almost all the final states unavailable. 

The slight variation in the energy and in the absolute value of the velocity observed with the FFMC in Fig.\ref{mean_FFMC} is compatible with an overestimation of the Pauli principle which freezes the charge dynamics. 

In the following subsections, we discuss the main results, in particular for the charge distribution and the mean energy and velocity, when both an initial Fermi-Dirac and a Maxwell-Boltzmann distribution are considered.

\subsection{Initial Maxwell-Boltzmann distribution}
To overcome the difficulty related to the frozen dynamics due to the Fermi-Dirac distribution, Tadyszak et al. \cite{Tady} proposed to use a high temperature Maxwell-Boltzmann distribution as initial condition in place of the Fermi-Dirac one; the distribution temperature is heuristically set equal to $80 \, T$, $T$ being the room temperature, for the silicon case. Along these lines, we introduce the following Maxwell-Boltzmann distribution
\begin{equation}
	f(0,\mathbf k)=f_{MB}(\mathbf k)\equiv\exp\left(-\frac{\varepsilon-\mu}{k_B T^*}\right)\,,
\end{equation}
where the free parameters $\mu$ and $T^*$ are the electro-chemical potential and the temperature, respectively. They can be determined by equaling the charge densities calculated with $f_{FD}$ and $f_{MB}$: one of the two parameters may be fixed and the other one determined by charge equality.. 
In particular, for each given Fermi level $\varepsilon_{F}$, one can fix $T$ at the room temperature $T^*=T$ and determine $\mu$, or set $\mu=0$ and calculate the temperature as $T^*=c \, T$. In table \ref{table4} the values of the constant $c$ and of the potential $\mu$ for different values of $\varepsilon_{F}$ are reported, as convenience the electron density $\rho$ for each value of $\varepsilon_{F}$.

\begin{table}[t]
	\caption{Values of the density $\rho$, coefficient $c$ and potential $\mu$ for different Fermi levels $\varepsilon_{F}$.	\label{table4}}
	\centering
	\begin{ruledtabular}
		\begin{tabular}{c|c|c|c|c}
			$\varepsilon_F \quad [\mbox{eV}]$ & $0.3$ & $0.4$ & $0.5$ & $0.6$\\
			\hline
			$\rho \quad [\mbox{$\mu$m$^{-2}$}]$ & $3.3867 \cdot 10^4$ & $5.9579 \cdot 10^4$ & $9.2638 \cdot 10^4$ & $1.3304 \cdot 10^5$\\
			\hline
			$c$ & $8.3082$ & $11.0197$ &$13.7410$ &$16.4672$\\
			\hline
			$\mu \quad \mbox{[eV]}$ & $0.1094$ & $0.1240$ &$0.1354$ &$0.1448$\\
		\end{tabular}
	\end{ruledtabular}
\end{table}

\begin{figure}[h!]
	\centering
	\fbox{a)			\includegraphics[width=0.41\columnwidth]{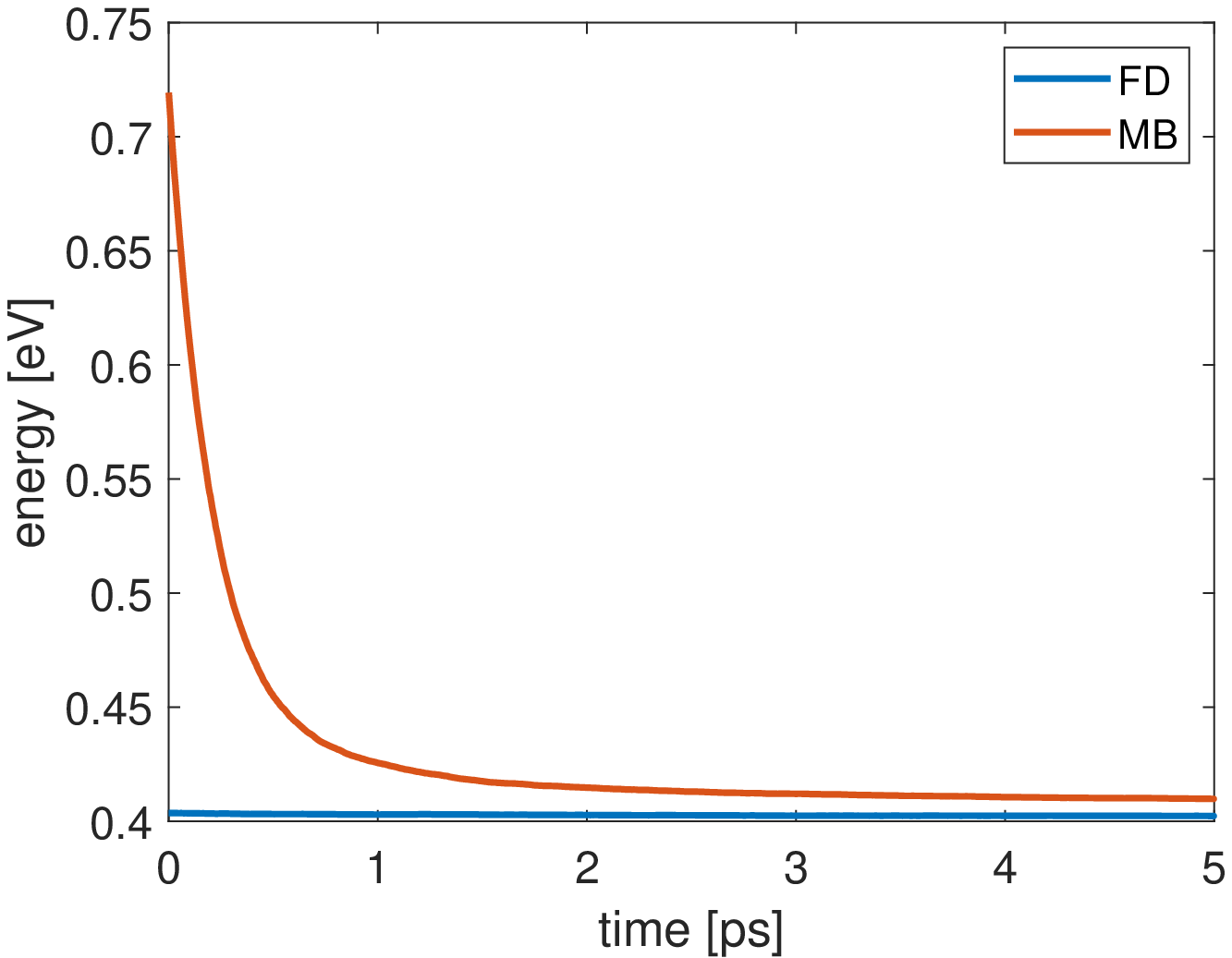}}
	\fbox{b)			\includegraphics[width=0.41\columnwidth]{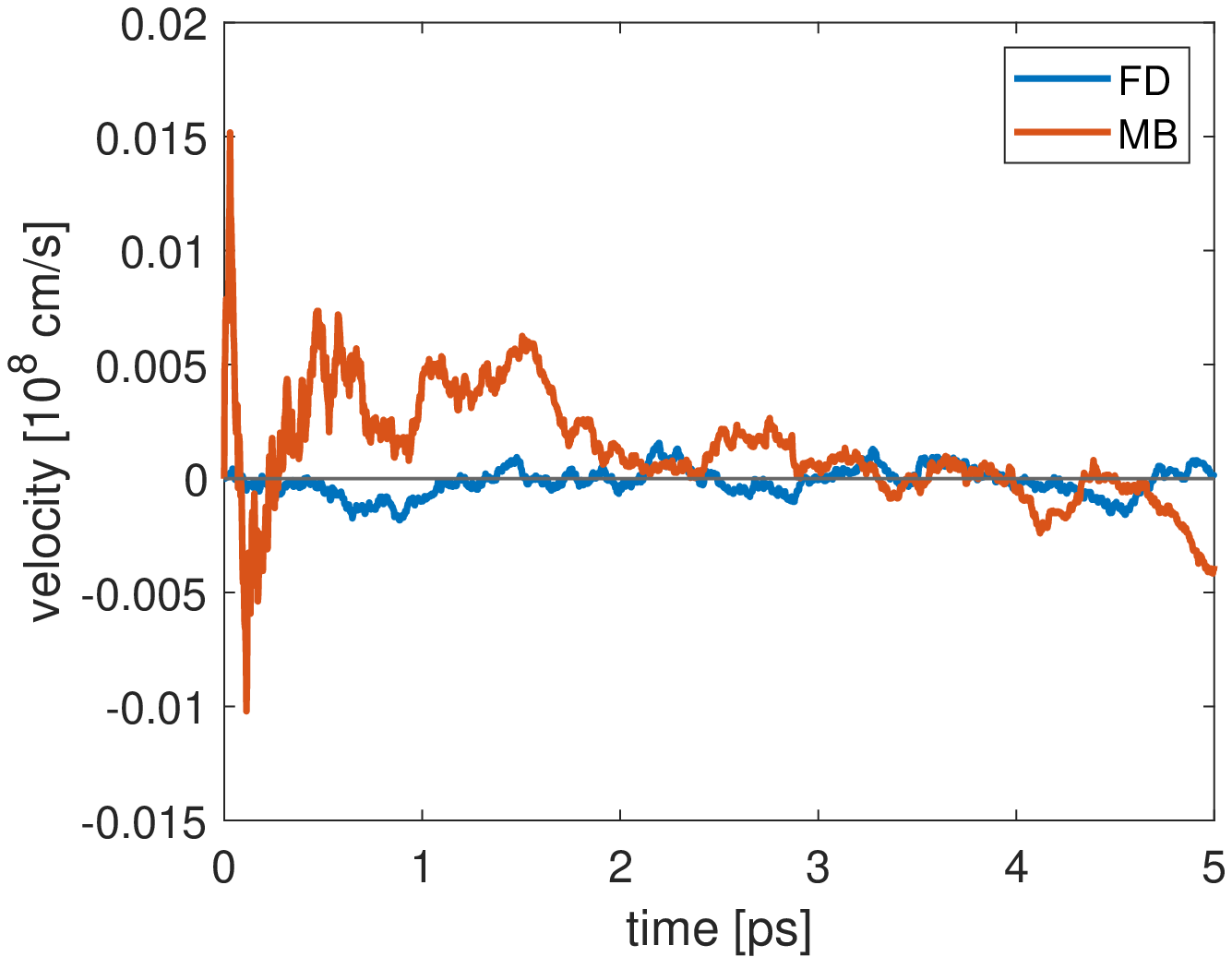}}\\
	\caption{Mean energy a) and mean velocity b) in the FFMC case when $\varepsilon_{F}=0.6$ eV and $E=0$ kV/cm. \label{0V_FFMC_en}}
\end{figure}

\begin{figure}[h!]
	\centering
	\fbox{a)			\includegraphics[width=0.41\columnwidth]{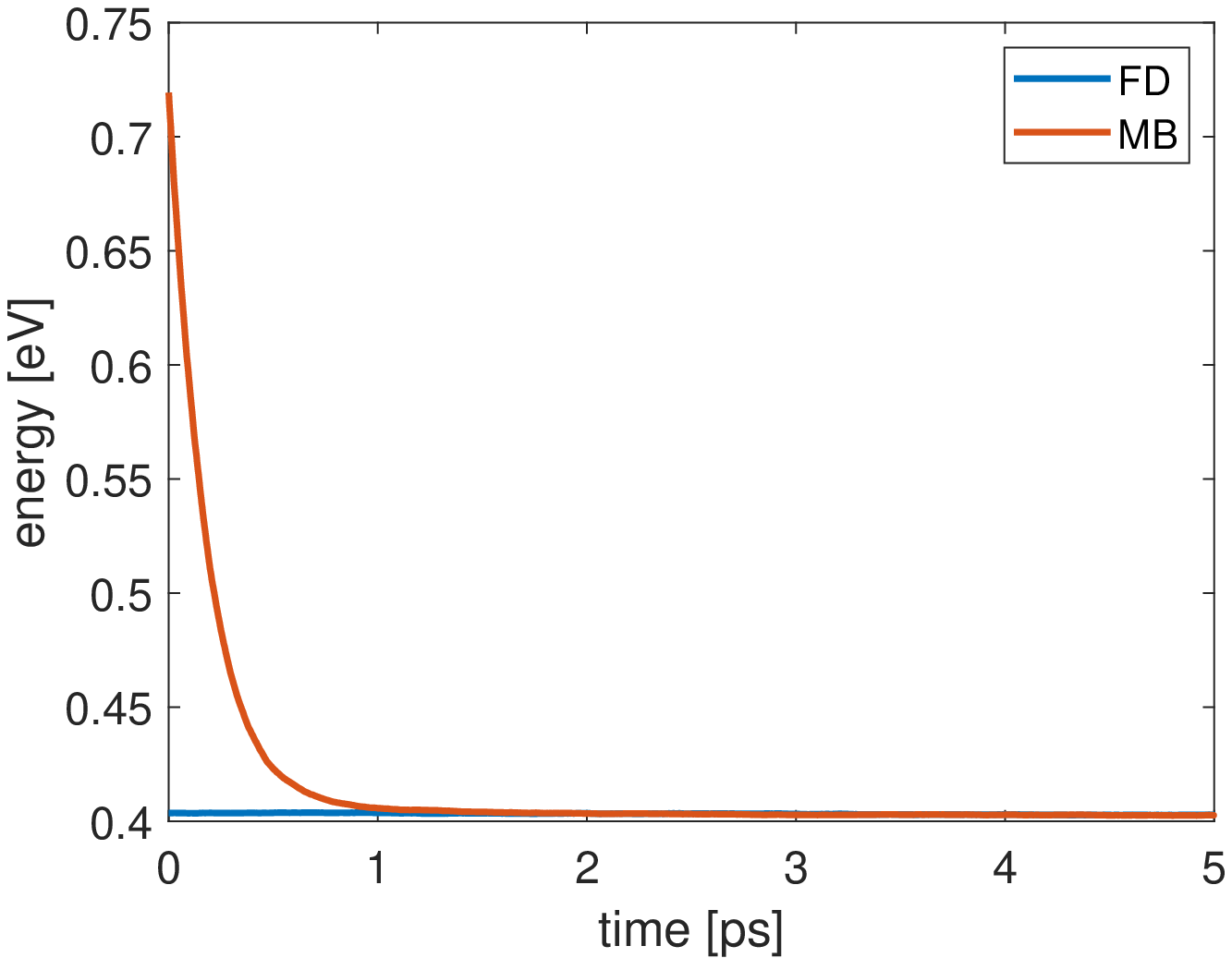}}
	\fbox{b)			\includegraphics[width=0.41\columnwidth]{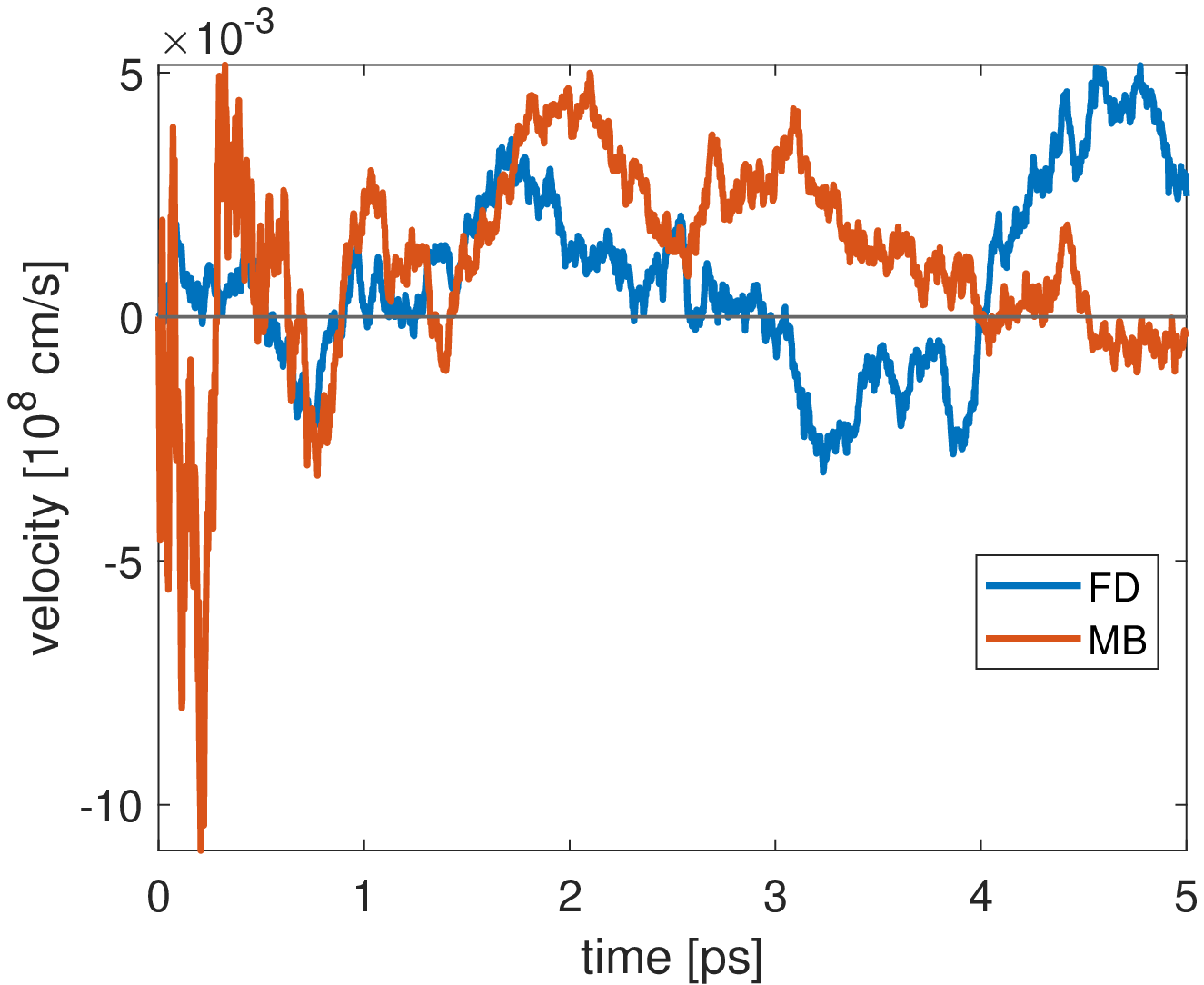}}\\
	\caption{Mean energy a) and mean velocity b) in the NEMC case when $\varepsilon_{F}=0.6$ and $E=0$ kV/cm. \label{MB_DSMC_en}}
\end{figure}

\begin{figure}[h!]
	\centering
	\fbox{a)			\includegraphics[width=0.41\columnwidth]{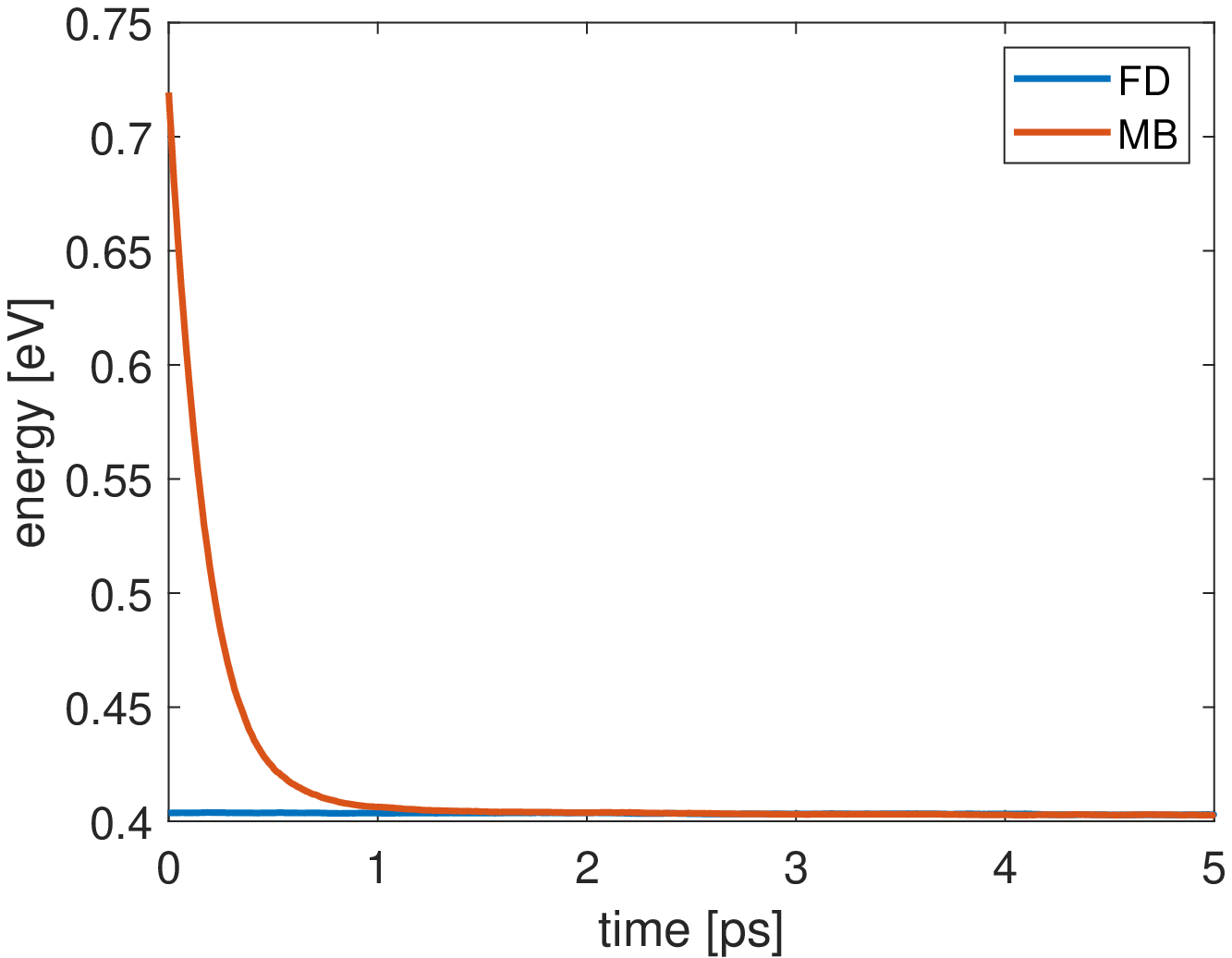}}
	\fbox{b)			\includegraphics[width=0.41\columnwidth]{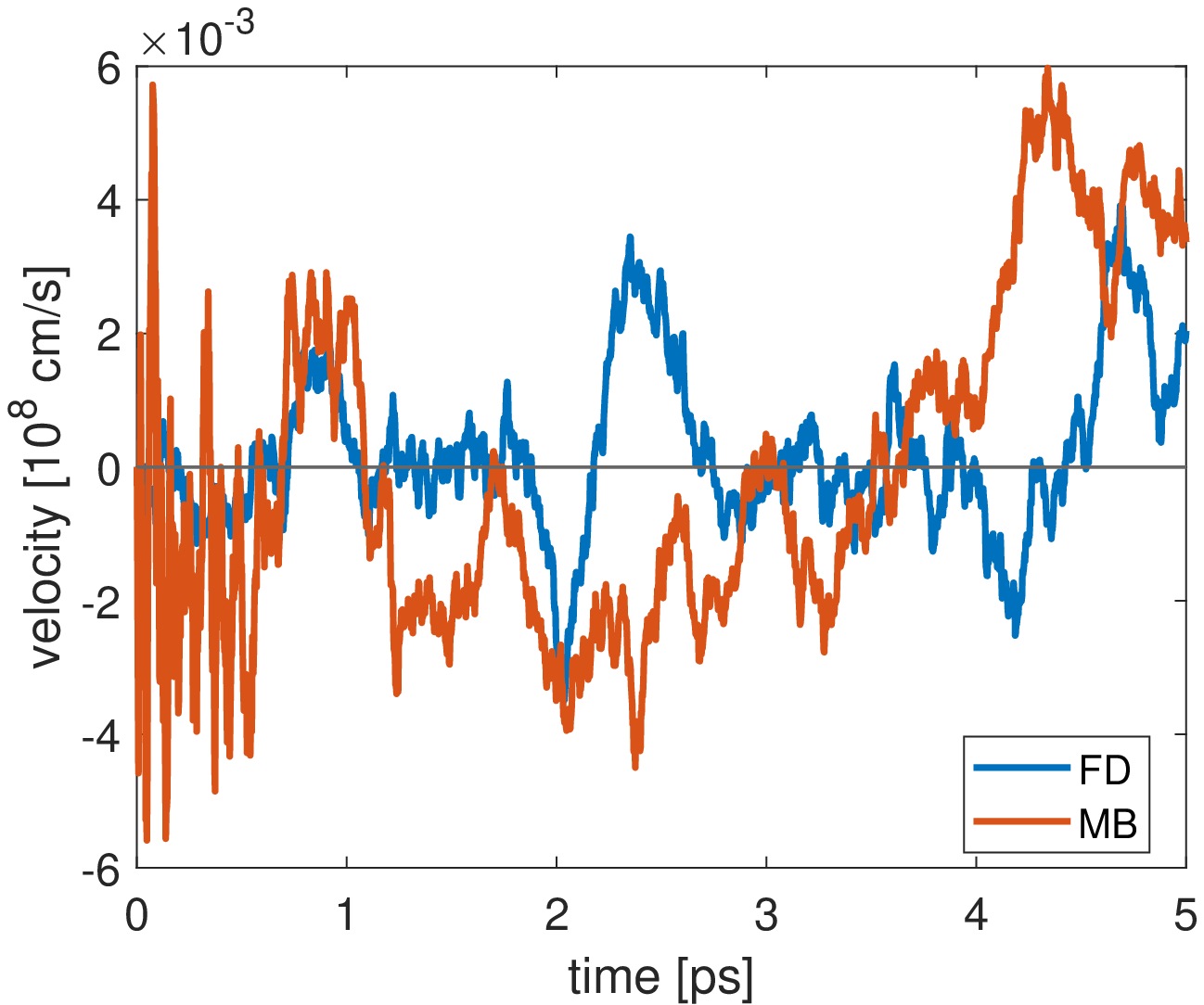}}\\
	\caption{Mean energy a) and mean velocity in the EMC case when $\varepsilon_{F}=0.6$ eV and $E=0$ kV/cm \label{EMC_en}}
\end{figure}

In the absence of an applied electric field, the mean energy and velocity, calculated either when the Fermi-Dirac or the Maxwell-Boltzmann distribution are taken as initial condition, have to converge to the same stationary values. This convergence is reached for all three Monte Carlo methods, NEMC, SEMC and FFMC, respectively, for different values of the Fermi level $\varepsilon_{F}$. In Fig. \ref{0V_FFMC_en} the average energy and velocity are shown when the Fermi-Dirac (FD) and the Maxwell-Boltzmann (MB) initial conditions are imposed in the FFMC with $\varepsilon_{F}=0.6$ eV. 
The same behavior is present when the NEMC and the SEMC are used (see Figs. \ref{MB_DSMC_en}-\ref{EMC_en}).

\begin{figure}[h!]
	\centering
	\fbox{a)			\includegraphics[width=0.41\columnwidth]{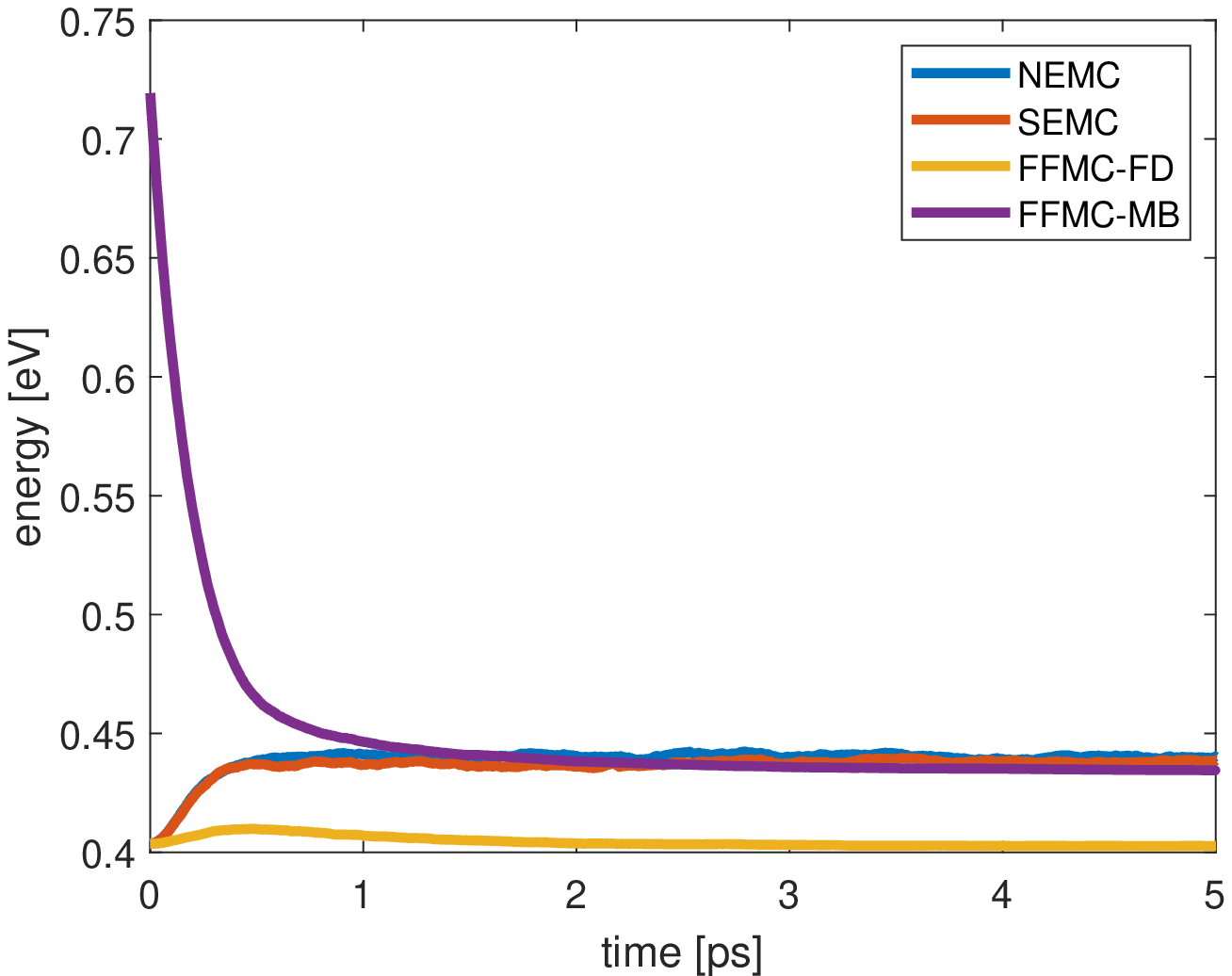}}
	\fbox{b)			\includegraphics[width=0.41\columnwidth]{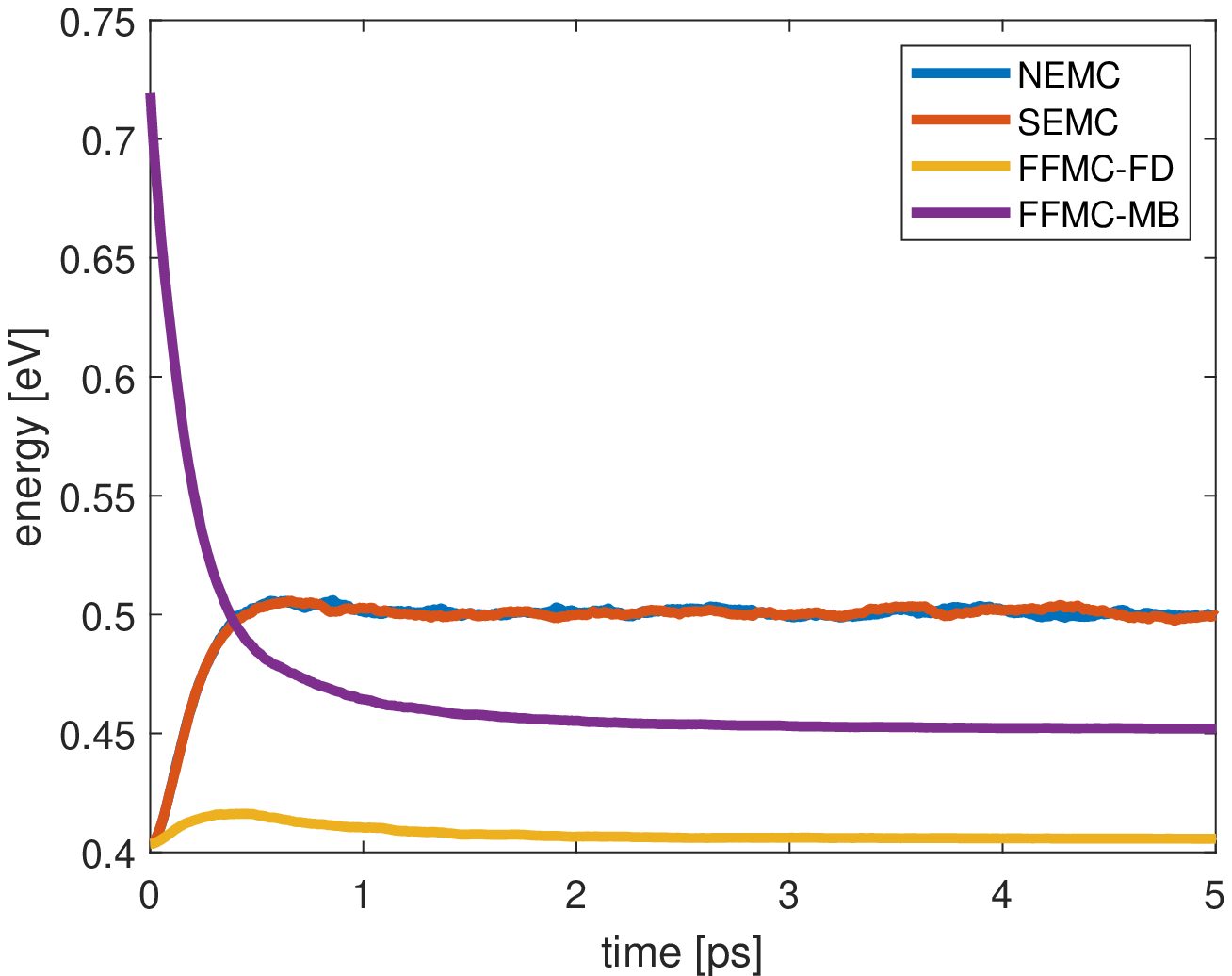}}\\
	\caption{Mean energy in the NEMC, SEMC, FFMC, with FD and MB initial distributions, cases when $\varepsilon_{F}=0.6$ eV and $E=10$ kV/cm, a), $\varepsilon_{F}=0.6$ eV and $E=20$ kV/cm, b).   \label{cp_01_en}}
\end{figure}

\begin{figure}[h!]
	\fbox{a)			\includegraphics[width=0.41\columnwidth]{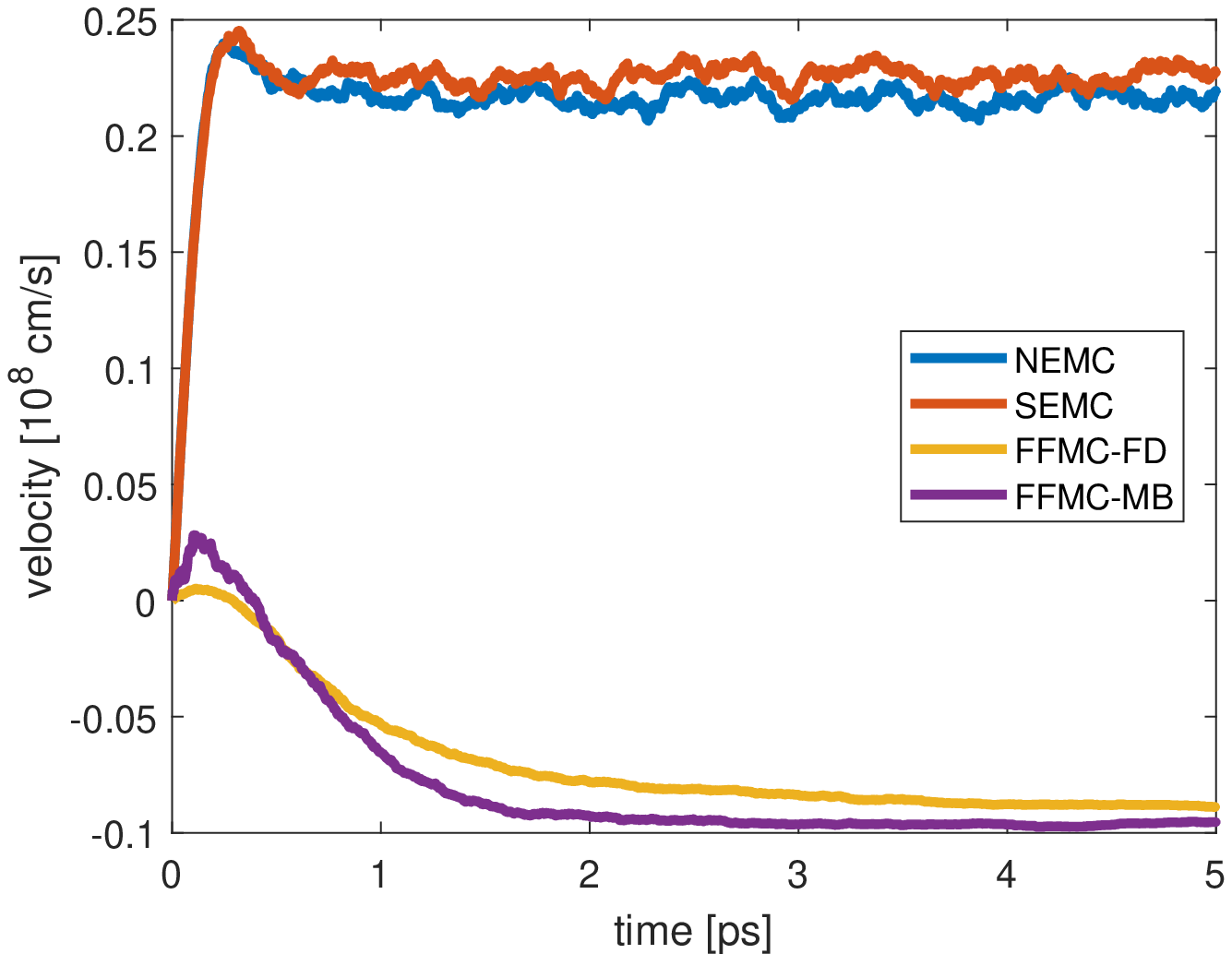}}
	\fbox{b)			\includegraphics[width=0.41\columnwidth]{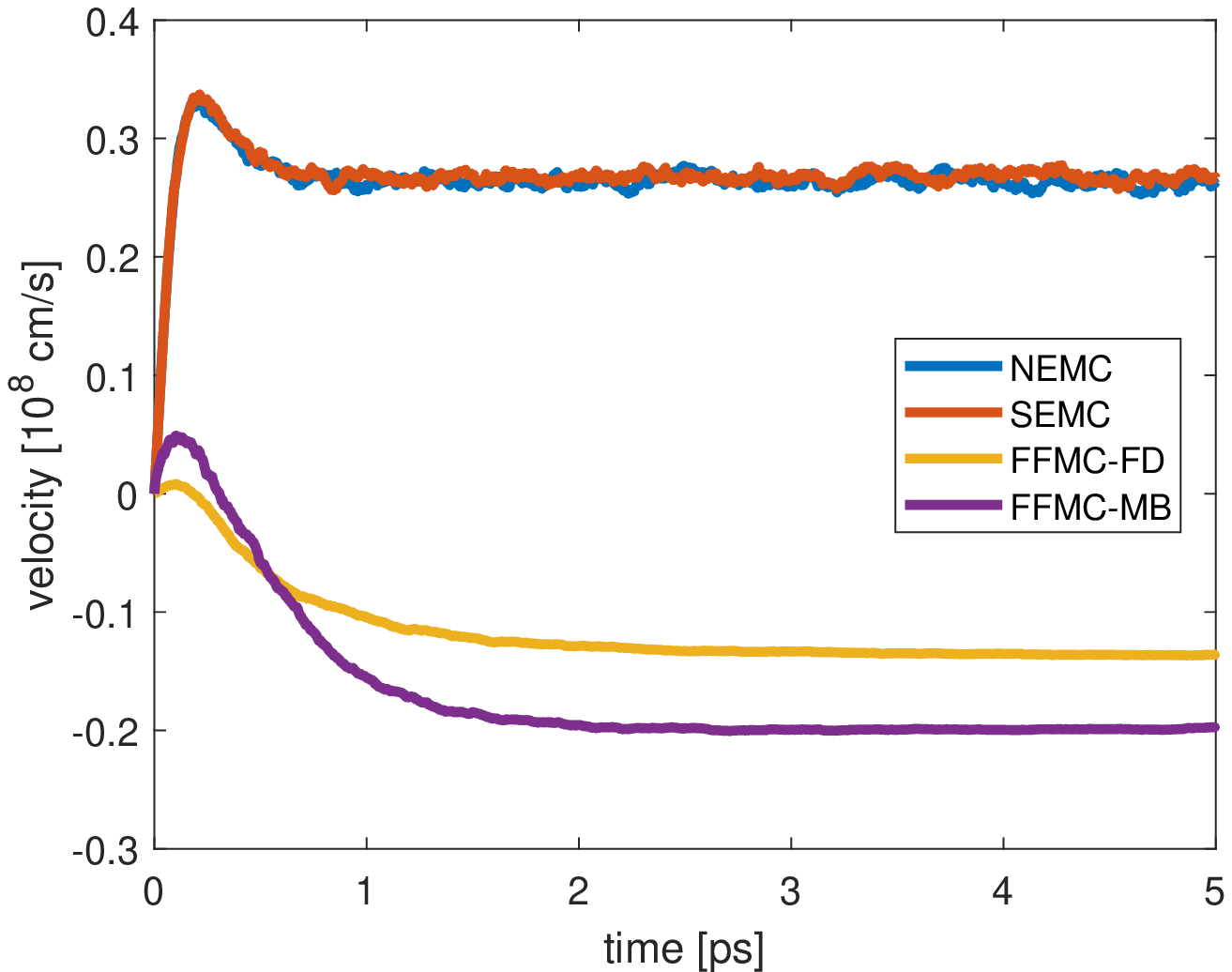}}\\
	\caption{Mean velocity in the NEMC, SEMC, FFMC, with FD and MB initial distributions, cases when $\varepsilon_{F}=0.6$ eV and $E=10$ kV/cm, a), $\varepsilon_{F}=0.6$ eV and $E=20$ kV/cm, b).   \label{cp_01_vel}}
\end{figure}

In the presence of an applied electric field $E$, the mean energy and velocity calculated with the SEMC and the NEMC approaches are in good agreement, while they are totally different with the FFMC. They are shown in Figs. \ref{cp_01_en} and \ref{cp_01_vel}, respectively, for $\varepsilon_{F}=0.6$ eV and $E=10$ and $20$ kV/cm. The energy given by the FFMC with an initial Fermi Dirac (FFMC-FD) distribution remains about constant, while with an initial Maxwell-Boltzmann distribution (FFMC-MB) it has an initial peak followed by a fast decrease; at $5$ ps the values obtained with the FD and MB distribution have a difference of about $9$ \%.

As it is evident from Fig. \ref{cp_01_vel}, also by using an initial Maxwell-Boltzmann distribution, the velocity has the same behavior as in Fig. \ref{mean_FFMC} b), with an initial small rising portion followed by a descent towards negative values, always higher in absolute value with respect to those obtained by considering the Fermi-Dirac distribution. The difference is considerable, of about $50$ \% for high values of the Fermi energy and of the applied electric fields. The choice of an initial high temperature Maxwell-Boltzmann distribution in place of a Fermi-Dirac produces a greater possibility of movement of the particles because there are fewer fully occupied states and the overestimation of the effect of applying the Pauli principle also at the end of each free flight is smaller, as noted in \cite{Tady} as well. Therefore, the absolute value of the velocity is appreciably higher.

\begin{figure}[h!]
	\centering
	\fbox{a)			\includegraphics[width=0.41\columnwidth]{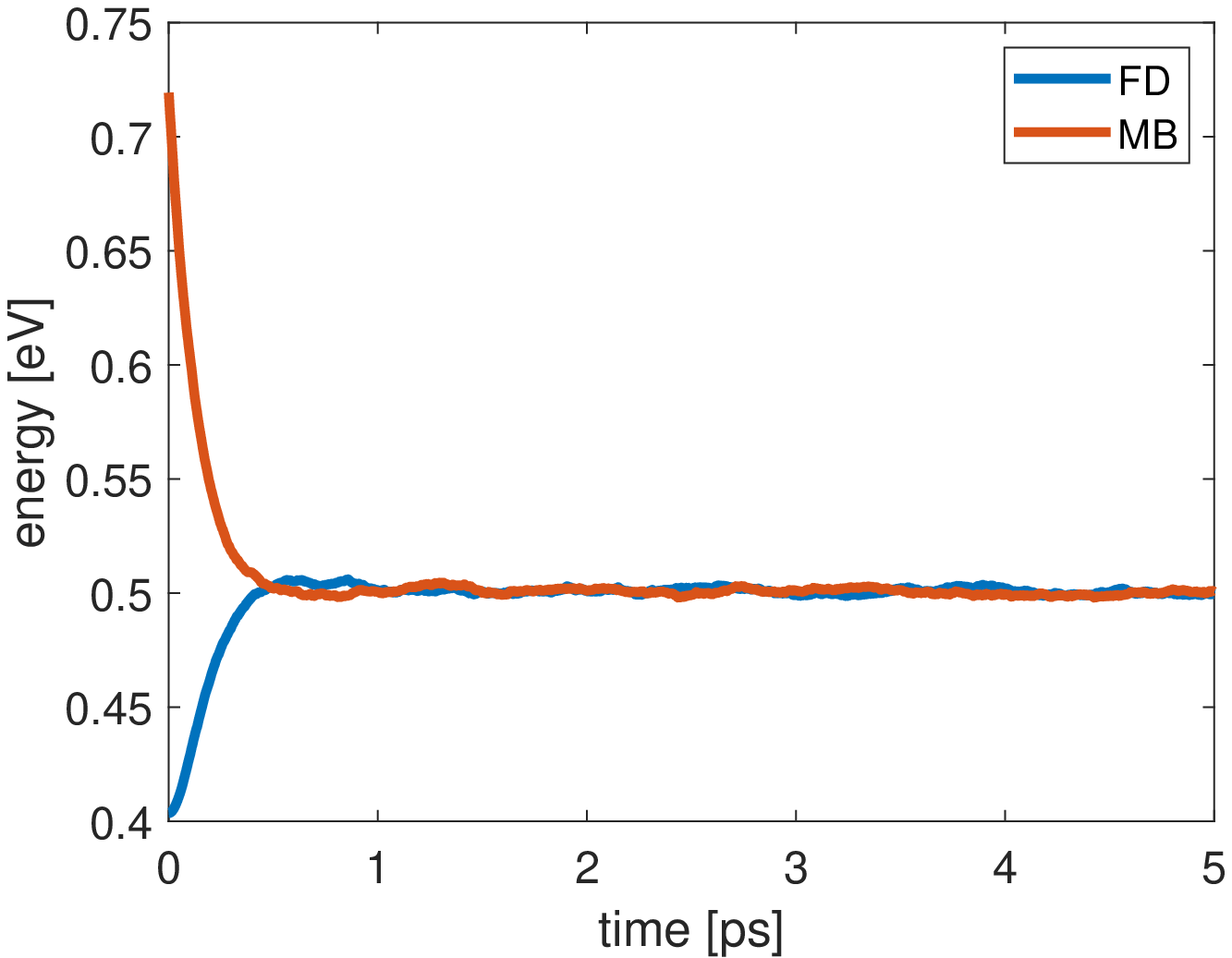}}
	\fbox{b)			\includegraphics[width=0.41\columnwidth]{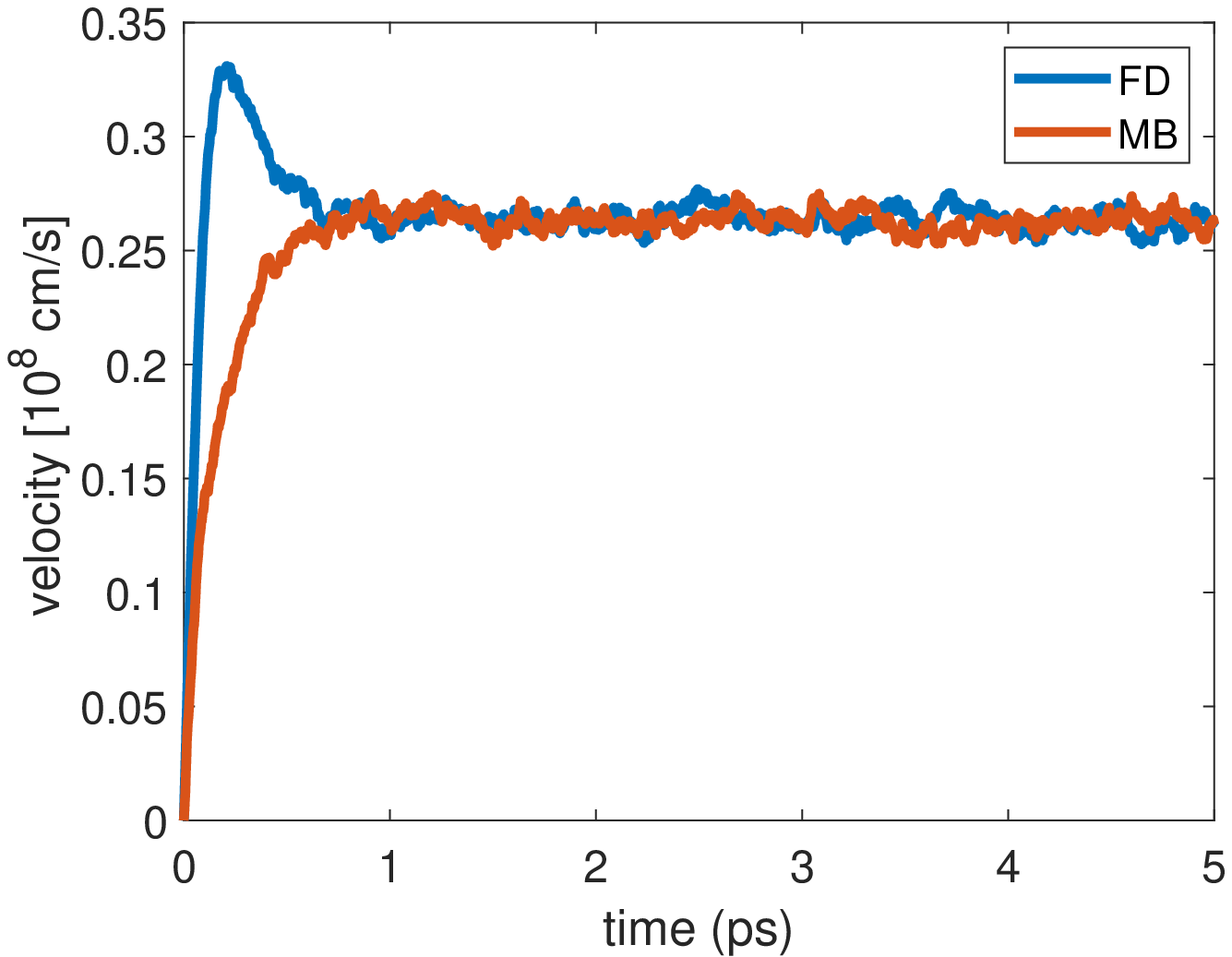}}\\
	\fbox{c)			\includegraphics[width=0.41\columnwidth]{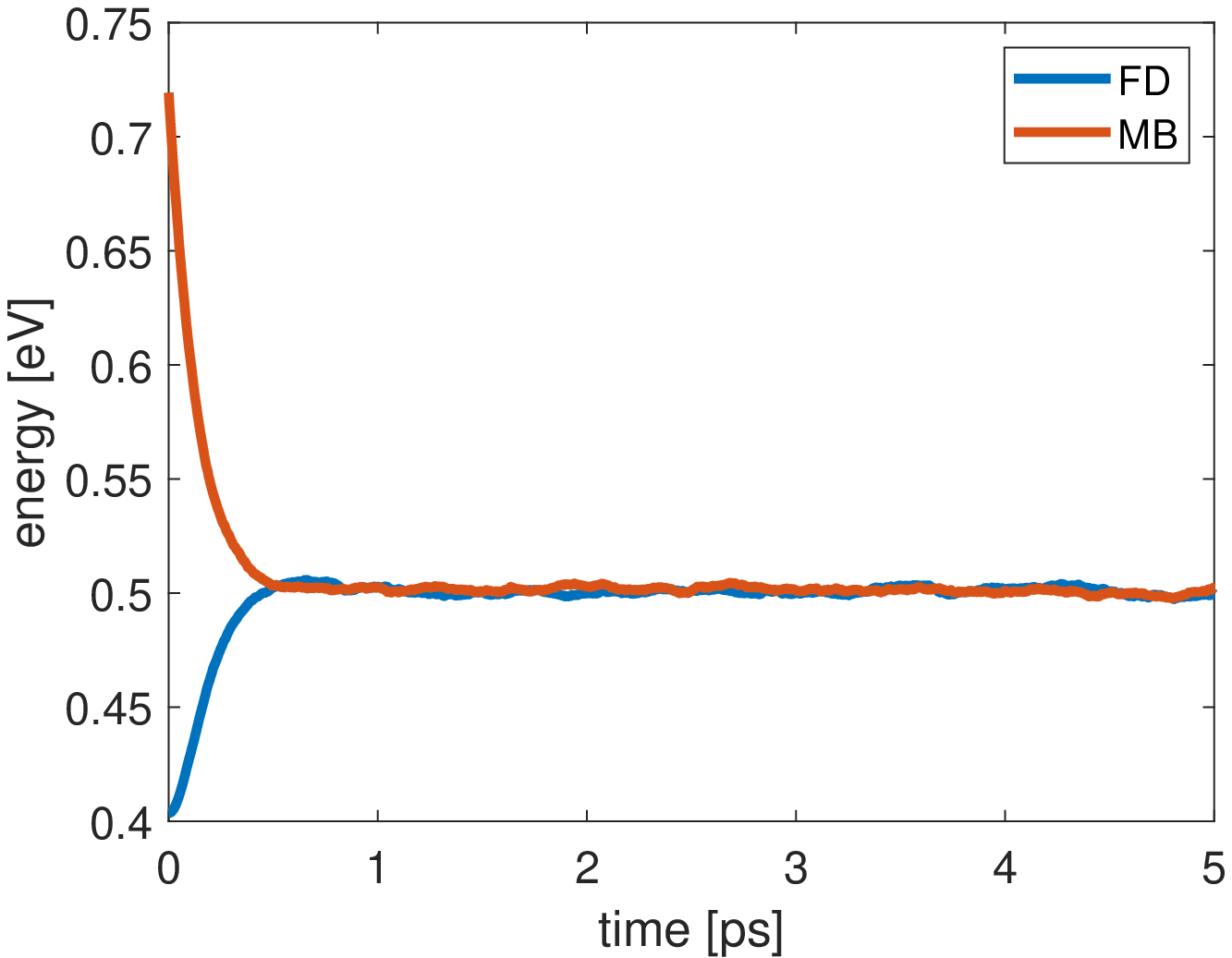}}
	\fbox{d)			\includegraphics[width=0.41\columnwidth]{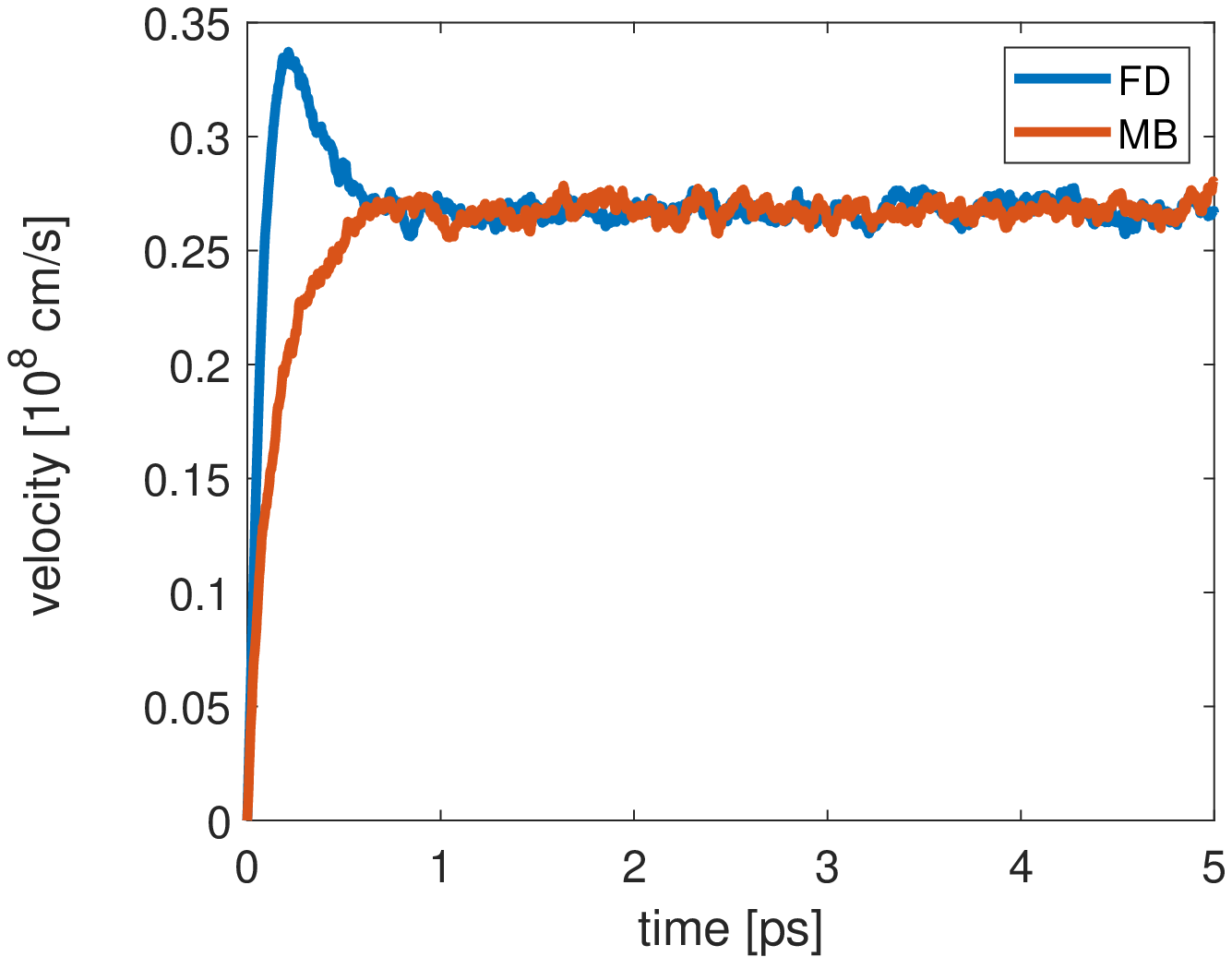}}\\
	\caption{Mean energy and velocity in the NEMC, a) and b), and in the SEMC, c) and d), cases when the FD or the MB initial distribution are used; $\varepsilon_{F}=0.6$ eV and $E=20$ kV/cm.  \label{cp_DSMC_EMC_FD_MB}}
\end{figure}

The SEMC and NEMC simulations are not affected by the choice of the initial condition; the mean energy and velocity reach the same stationary values; the only difference is an initial overshoot in the velocity that is not present when the initial Maxwell-Boltzmann distribution is used (see Fig. \ref{cp_DSMC_EMC_FD_MB}).

\subsection{The charge distribution of the FFMC}
The results obtained by using the FFMC procedure, i.e. almost constant mean energies and negative mean velocities, are certainly unphysical and not in accordance with the solution of the Boltzmann equation, as well as with the SEMC and NEMC methods, and deserve a deeper analysis.

\begin{figure}[h!!]
	\centering
	\fbox{a)			\includegraphics[width=0.41\columnwidth]{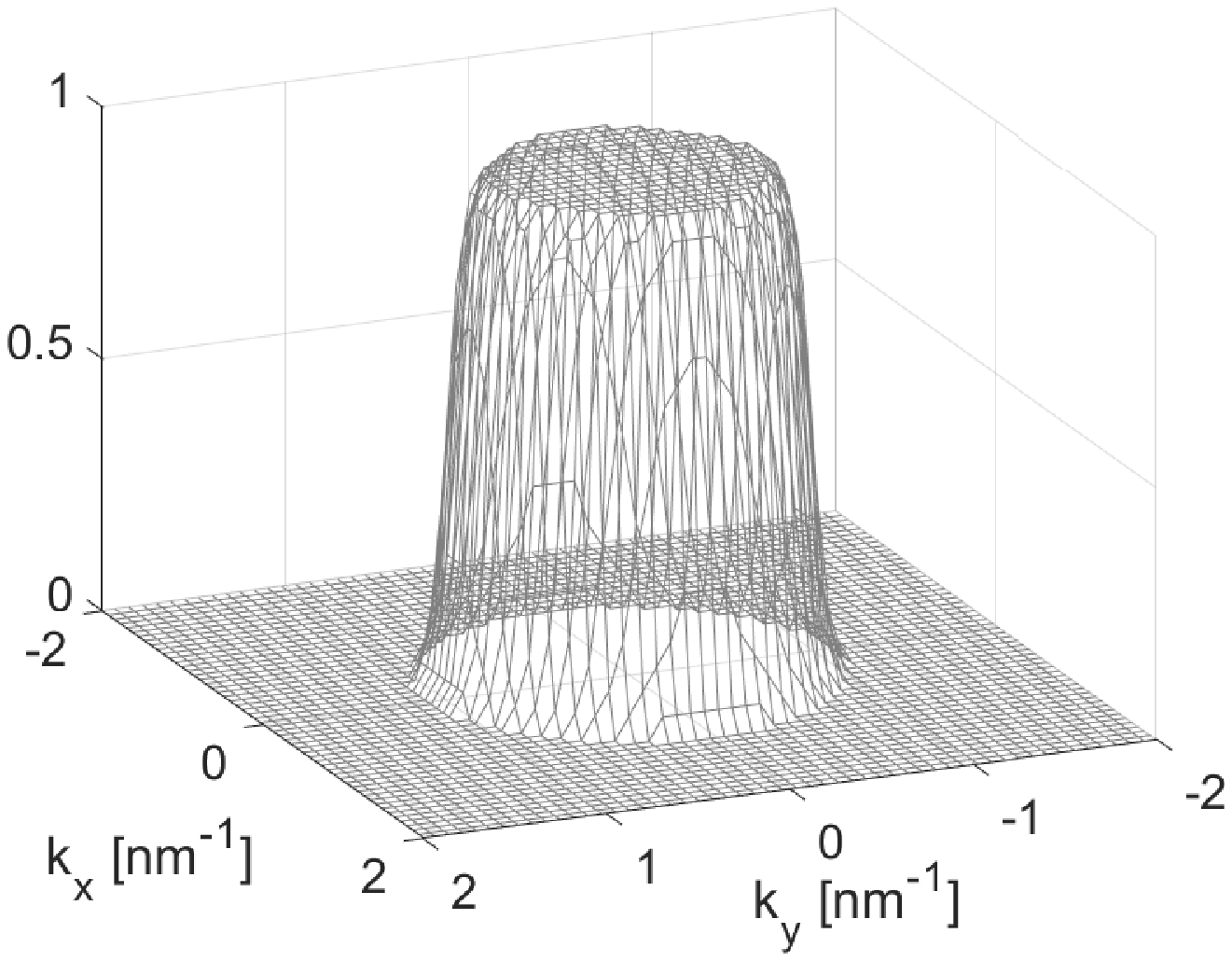}}
	\fbox{c)			\includegraphics[width=0.41\columnwidth]{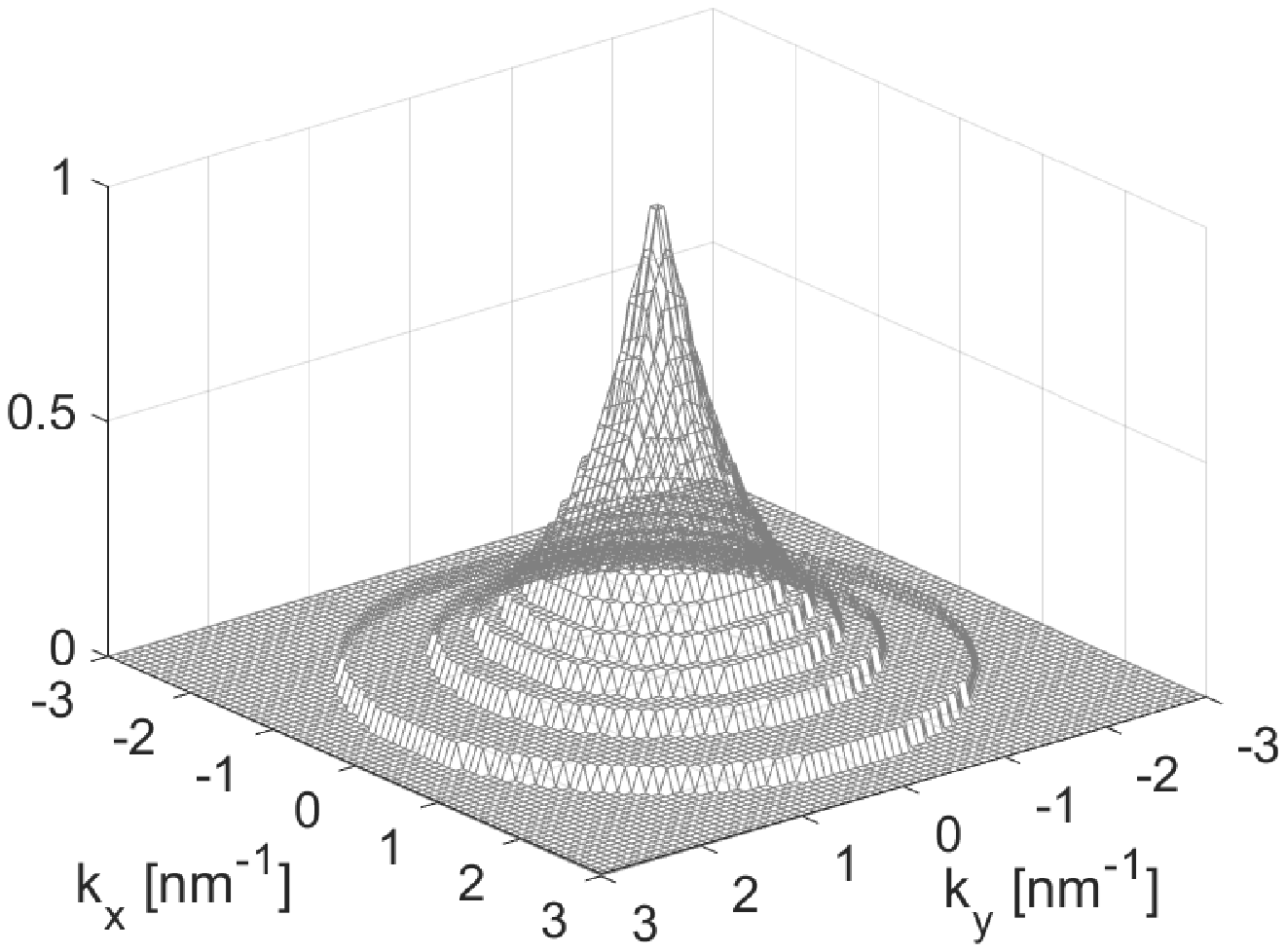}}\\
	\fbox{b)			\includegraphics[width=0.41\columnwidth]{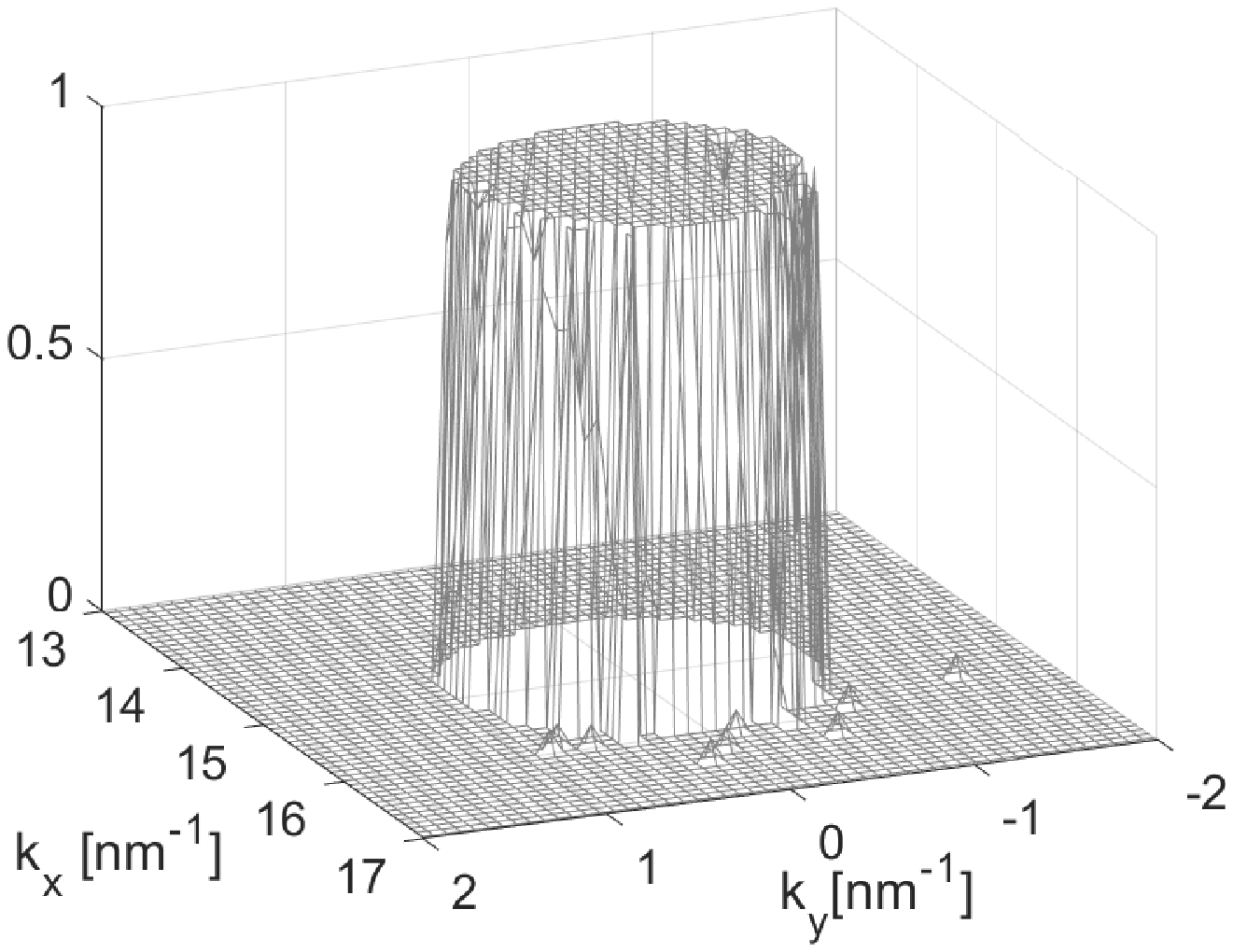}}
	\fbox{d)			\includegraphics[width=0.41\columnwidth]{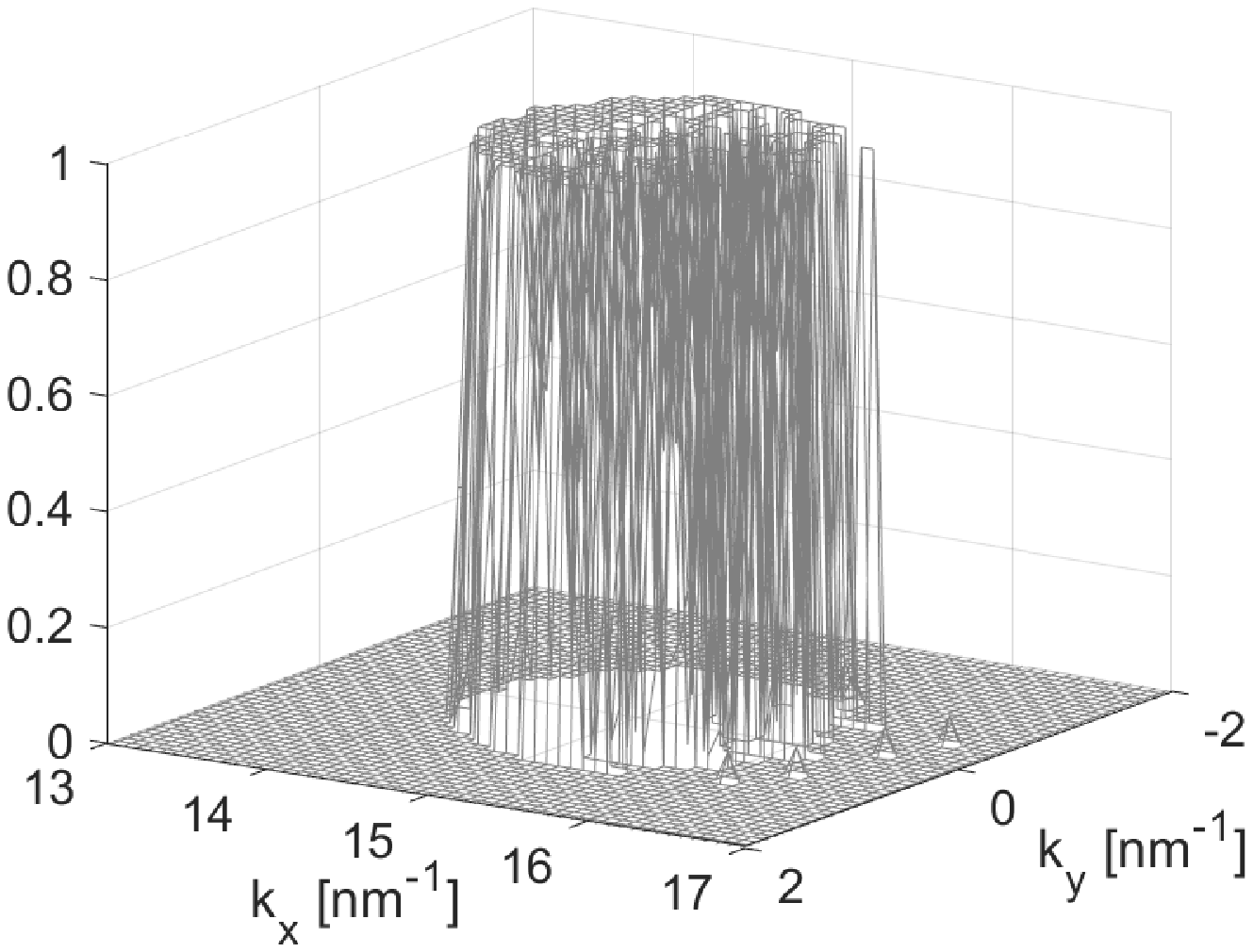}}\\
	\caption{FFMC scheme. Initial Fermi-Dirac a) and charge distribution at $5$ ps b). Initial Maxwell-Boltzmann c) and charge distribution at $5$ ps d). $\varepsilon_{F}=0.6$ eV and $E=20$ kV/cm.  \label{cp_distr_06_2V}}
\end{figure}

The comparison between the charge distributions obtained in the FFMC simulation when the Fermi-Dirac or the Maxwell-Boltzmann distribution are taken as initial conditions is shown in Fig. \ref{cp_distr_06_2V} for $\varepsilon_{F}=0.6$ eV and $E=20$ kV/cm. 
The initial Fermi-Dirac distribution undergoes only negligible changes during the time evolution (see Figs. \ref{cp_distr_06_2V} a,b); the Maxwell-Boltzmann as initial distribution allows a coarser collocation of the particles with high mean energy because it occupies a portion of the $\bf k$-space twice larger than that of the initial Fermi-Dirac (see Fig. \ref{cp_distr_06_2V} c). The initial MB distribution does not seem to evolve towards a realistic particle distribution but, as time goes, it becomes more similar to an irregular FD-like distribution, as clearly shown in Fig. \ref{cp_distr_06_2V} (d). Small voids are present mainly in the front of the distribution, due to the effect of the electric field, while the backside is more compact. It is evident that the initial Maxwell-Boltzmann conditions leads to a larger absolute velocity due to the greater possibility of movement of the charged particles. The same considerations hold when the density, i.e. the Fermi level $\varepsilon_{F}$, is lower; in this case the time necessary to reach a FD-like final distribution is longer. 

The dynamics generated by an initial FD distribution is trivial because almost all the final states result occupied during the simulation, on account of the Pauli principle imposed also at the end of each free flight. When the MB distribution is employed as initial condition, the dynamics becomes more complex but it seems to be rather a numerical effect than the simulation of a realistic physical situation described by the Boltzmann equations. The results are unphysical; the evolution of the distribution function is not physically satisfactory and the negative value of the mean velocity has no rationale at all.

\subsection{Negative mean velocity with the FFMC method}

\begin{figure}[h]
	\centering
	\fbox{a)			\includegraphics[width=0.41\columnwidth]{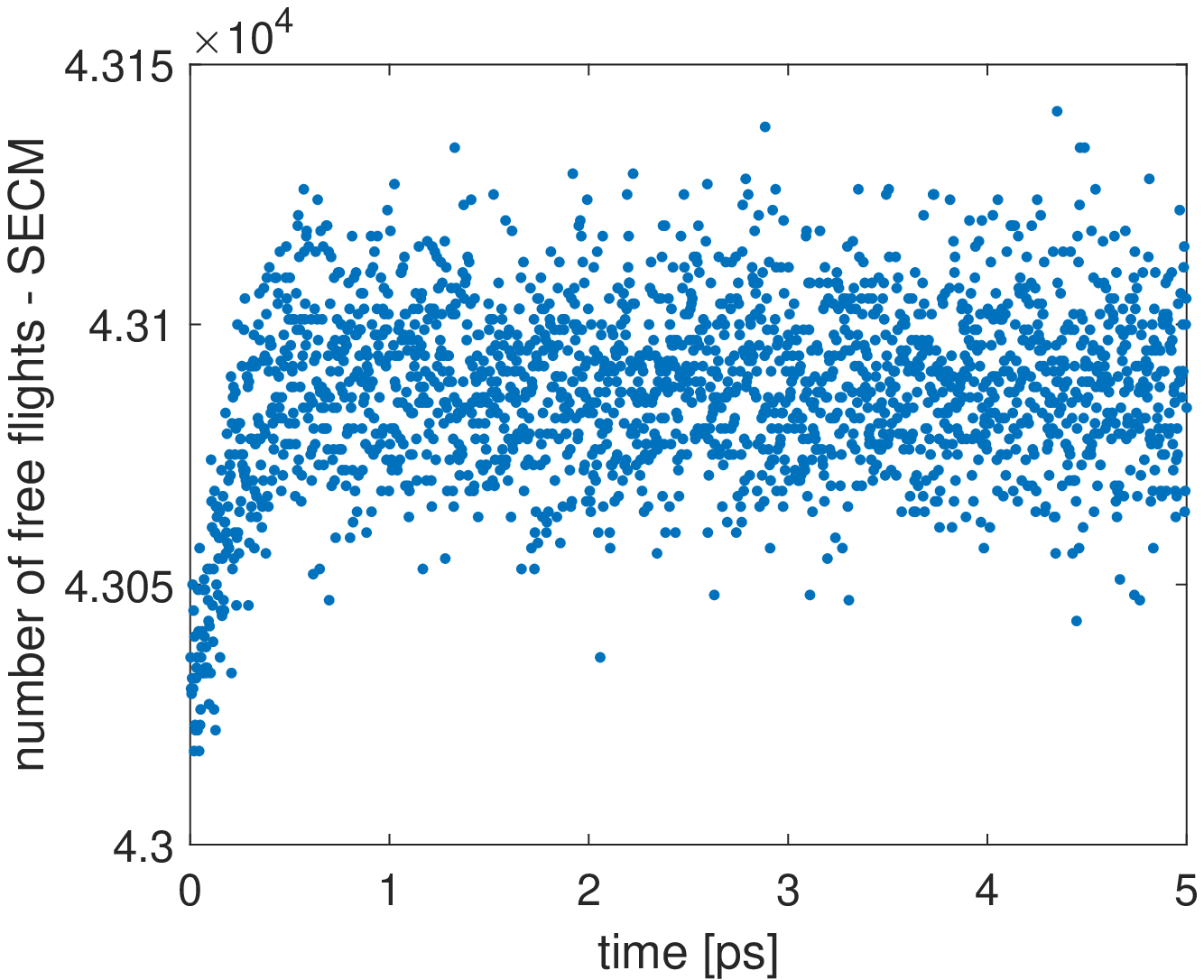}}
	\fbox{b)			\includegraphics[width=0.41\columnwidth]{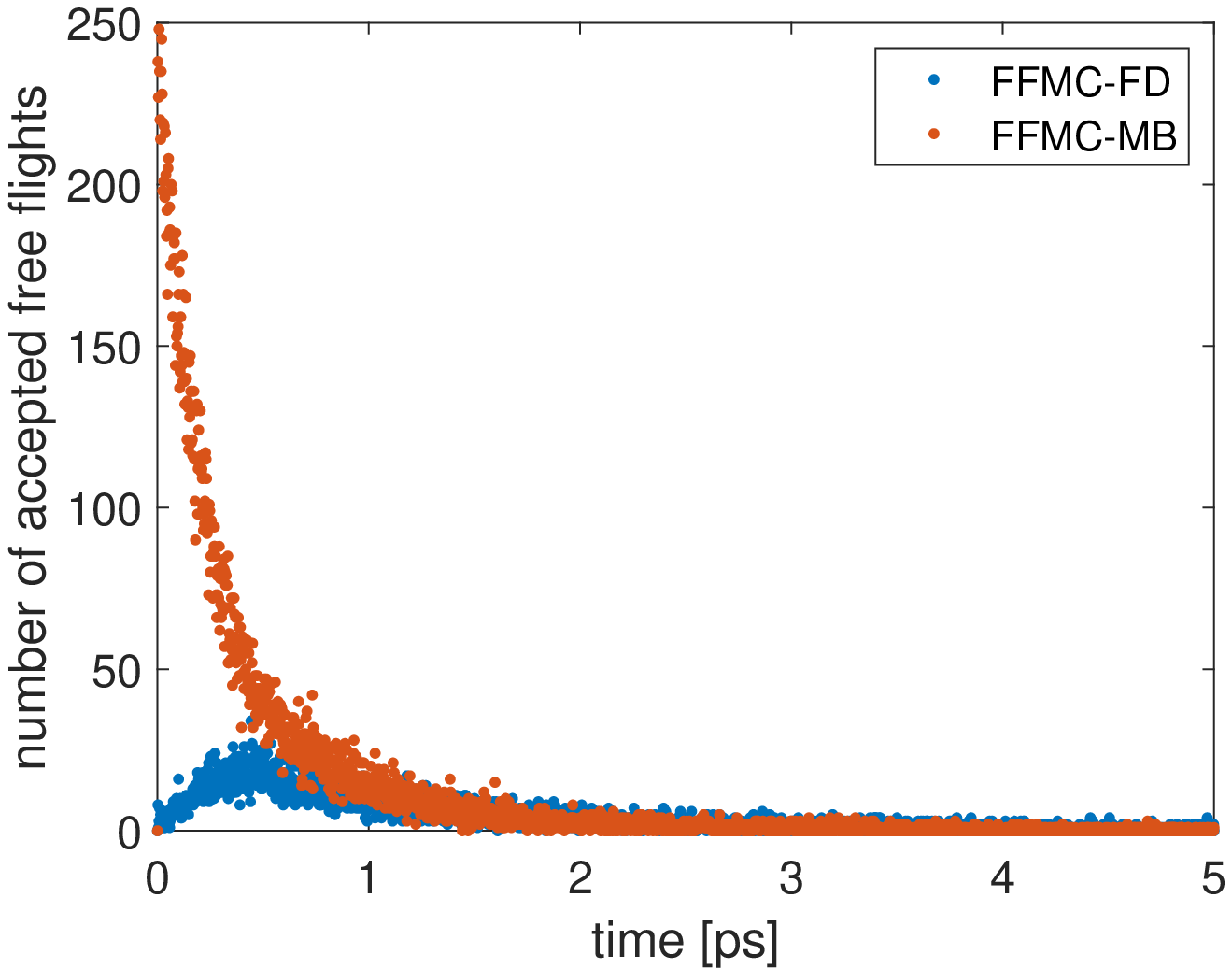}}\\
	\caption{Number of the accepted free flights for SEMC, with the initial FD, a), and for FFMC, b), when the FD and the MB are used, respectively.	\label{n_ff_acc}}
\end{figure}

To understand the origin of the negative mean velocity obtained with the FFMC approach, we investigate the main steps of the Monte Carlo procedure in the three methods, i.e. the free flights and the subsequent scattering events. In the NEMC procedure, the distribution function is translated as a whole at each time step $\Delta t$, and the number of the free flights is equal to that of the simulated particles $n_P=10^4$. In the SEMC scheme, the particles are followed one by one; this leads to an incorrect reconstruction of the distribution function, even if the number of free flights is of the same order as in the NEMC case, as shown in Fig. \ref{n_ff_acc} (a) for the case of an initial Fermi-Dirac distribution. In the NEMC approach, the Pauli principle is not imposed at the end of the free flight but only after the scatterings, so that all free flights are allowed and take place. If the Pauli principle is imposed also at the end of the free flights, the number of the accepted free flights is strongly reduced as reported in Fig. \ref{n_ff_acc} (b). Also when initially the MB distribution is used, the number of free flights is about of $250$ at early times but then rapidly decreases and becomes equal to the case with initial FD distribution. The use of the Pauli principle at the end of each free flight blocks the dynamics; it depletes the numerical sample and only few tens of events survive in comparison to the about $10^4$ in the other schemes. 

\begin{figure}[h]
	\centering
	\includegraphics[width=0.7\columnwidth]{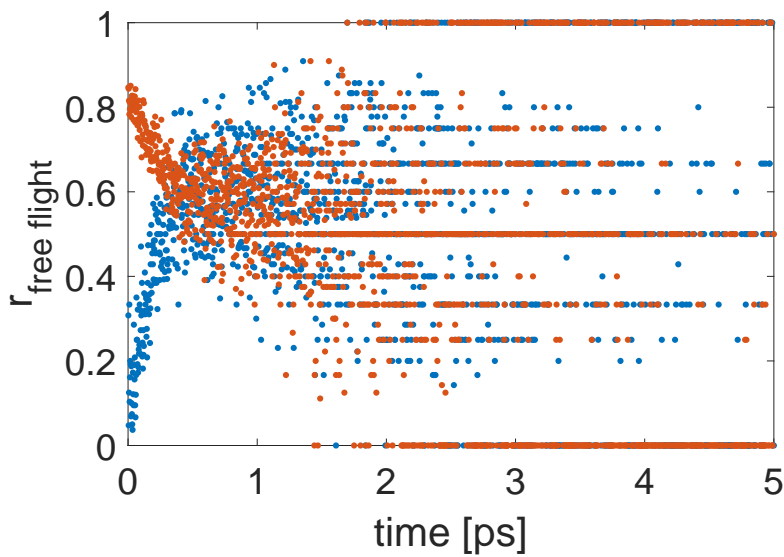}
	\caption{Ratio between the number of the accepted free flights with respect to their total number for the FFMC, with the FD, a), and the MB, b), initial condition, respectively.	\label{FFMC_ff_acc}}
\end{figure}

The ratio between the number of the accepted free flights with respect to the total is around $0.6$ until $t=1.5$ ps and drops very much afterwards, making the statistics not significant (see Fig. \ref{FFMC_ff_acc}).

\begin{figure}[h]
	\centering
	\includegraphics[width=1.0\columnwidth]{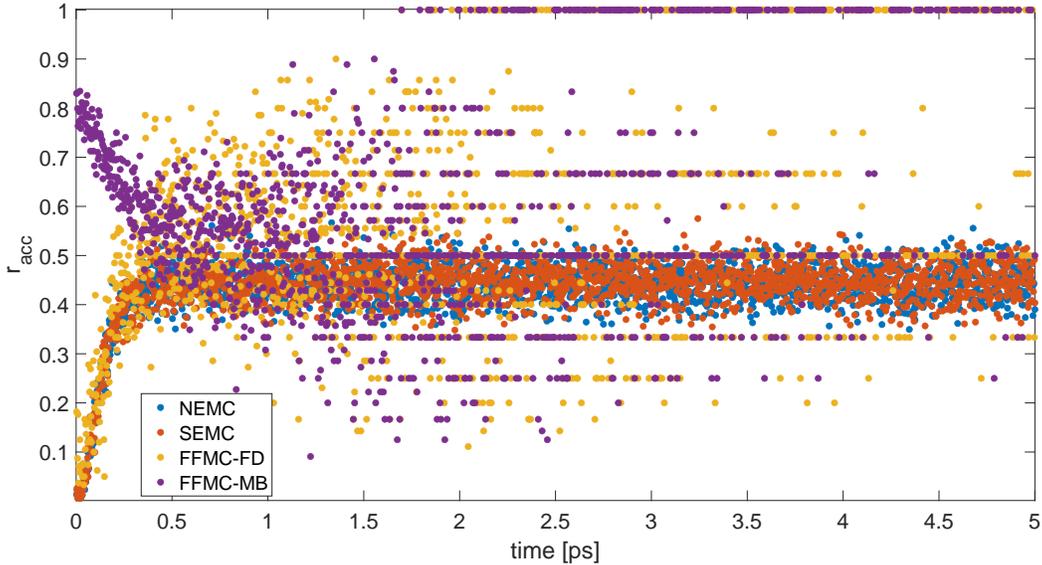}
	\caption{Ratio between the number of the accepted scatterings with respect to their total number with the FFMC, with the FD and the MB initial condition, respectively, and the SEMC and the NEMC schemes.	\label{sc_acc}}
\end{figure}

\begin{figure}[h]
	\centering
	\fbox{a)			\includegraphics[width=0.41\columnwidth]{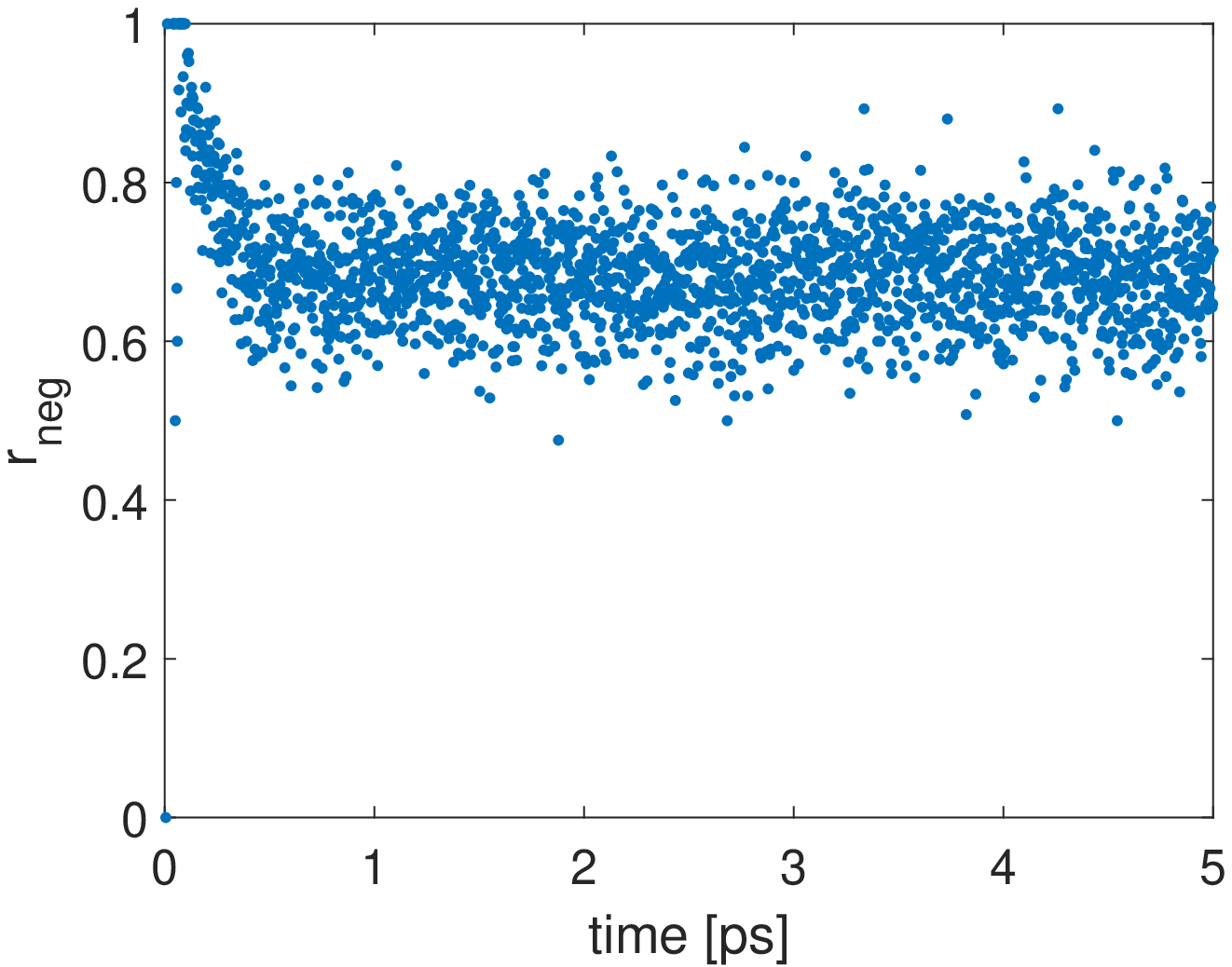}}
	\fbox{b)			\includegraphics[width=0.41\columnwidth]{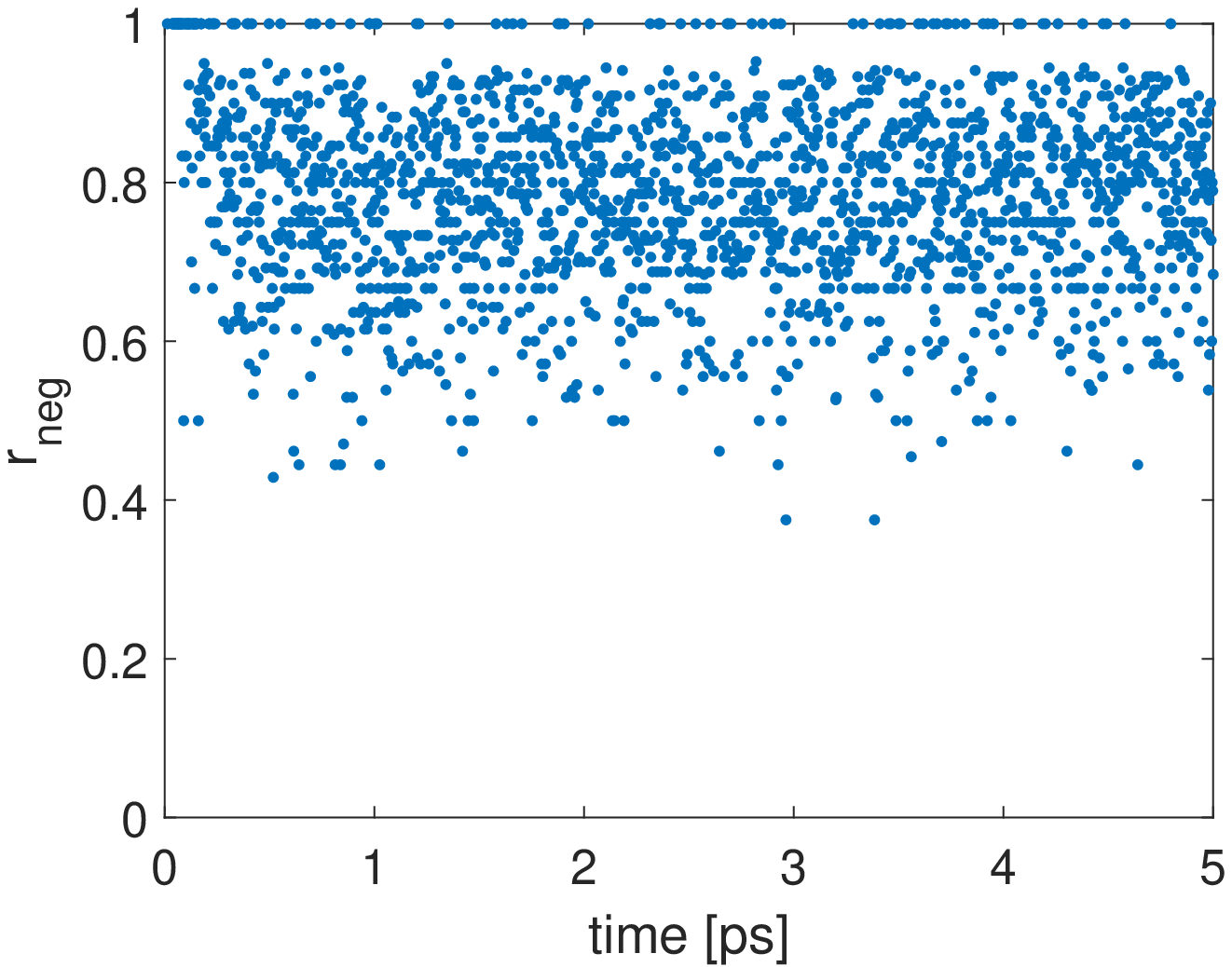}}\\
	\caption{Ratio between the number of emitted longitudinal optical a) and $K$ scatterings b), whose final velocity is negative with respect to their total number, for the NEMC approach.	\label{neg_LO}}
\end{figure}

The free flights, even a few ones, give a positive contribution to the mean velocity. Therefore, the negative values seen with the FFMC in Fig. \ref{cp_01_vel} should originate from the scattering events. The percentage of scatterings whose final states are available is shown in Fig. \ref{sc_acc}. 
For the NEMC and SEMC procedures the mean value is about $45$ \% with a standard deviation of about $4$ \%; in the FFMC case, we observe the same behavior shown in Fig. \ref{FFMC_ff_acc}, and after $1.5$ ps only very few non representative events are left. In order to understand the negative value of the mean velocity, we analyze all types of scatterings, in particular the fraction of those which lead to a final negative velocity. This fraction is reported in Fig. \ref{neg_LO} for the $LO$ and $K$ scatterings in the case of the NEMC (the SEMC scheme gives the same qualitative results) while the statistics is again not significant for the FFMC. Almost all the $LO$ and $K$ scatterings produce a negative final velocity in the first $0.5$ ps, then a mean constant value of about $70$ \%  with a small variance is reached for the $LO$ collisions (see Fig. \ref{neg_LO} a) and of about $80$ \% for the $K$ scatterings (see Fig. \ref{neg_LO} b).

\begin{figure}[h]
	\centering
	\fbox{a)			\includegraphics[width=0.41\columnwidth]{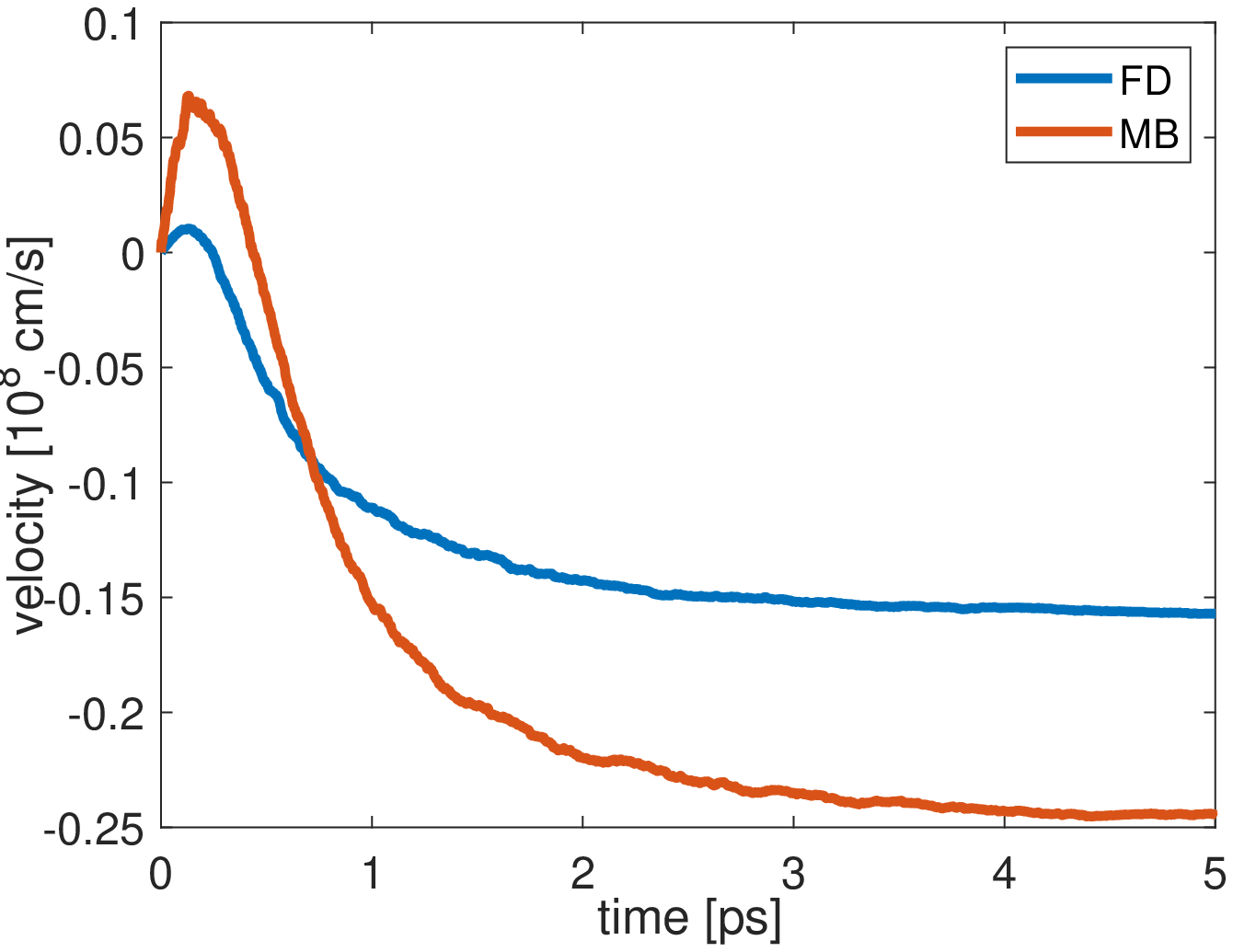}}
	\fbox{b)			\includegraphics[width=0.41\columnwidth]{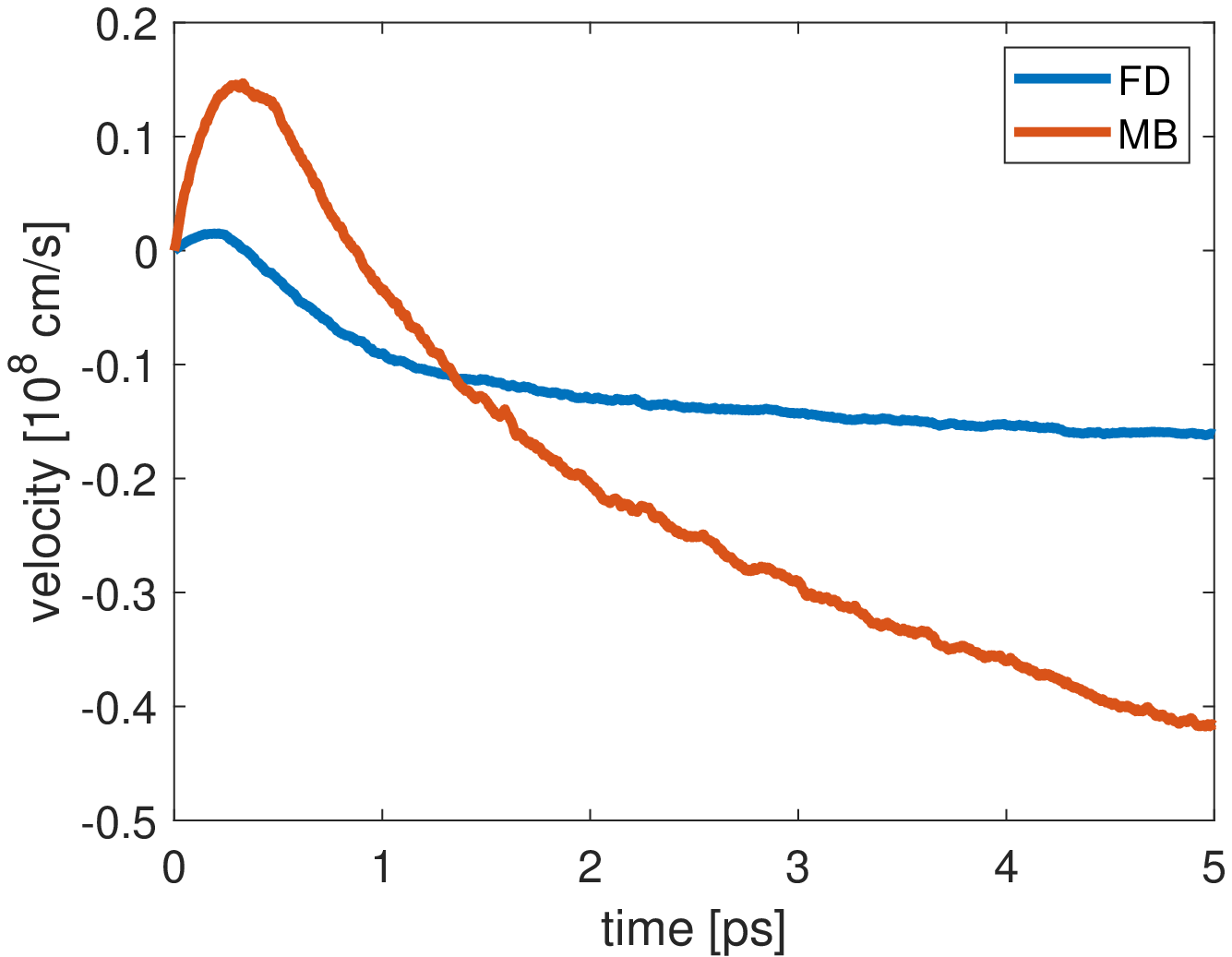}}\\
	\caption{Mean velocity when only the $LO$, a), and only $K$ scatterings, b), are included for the FFMC with the FD and the MB initial condition, respectively.	\label{LO_K_vel}}
\end{figure}

\begin{figure}[h]
	\centering
	\fbox{a)			\includegraphics[width=0.41\columnwidth]{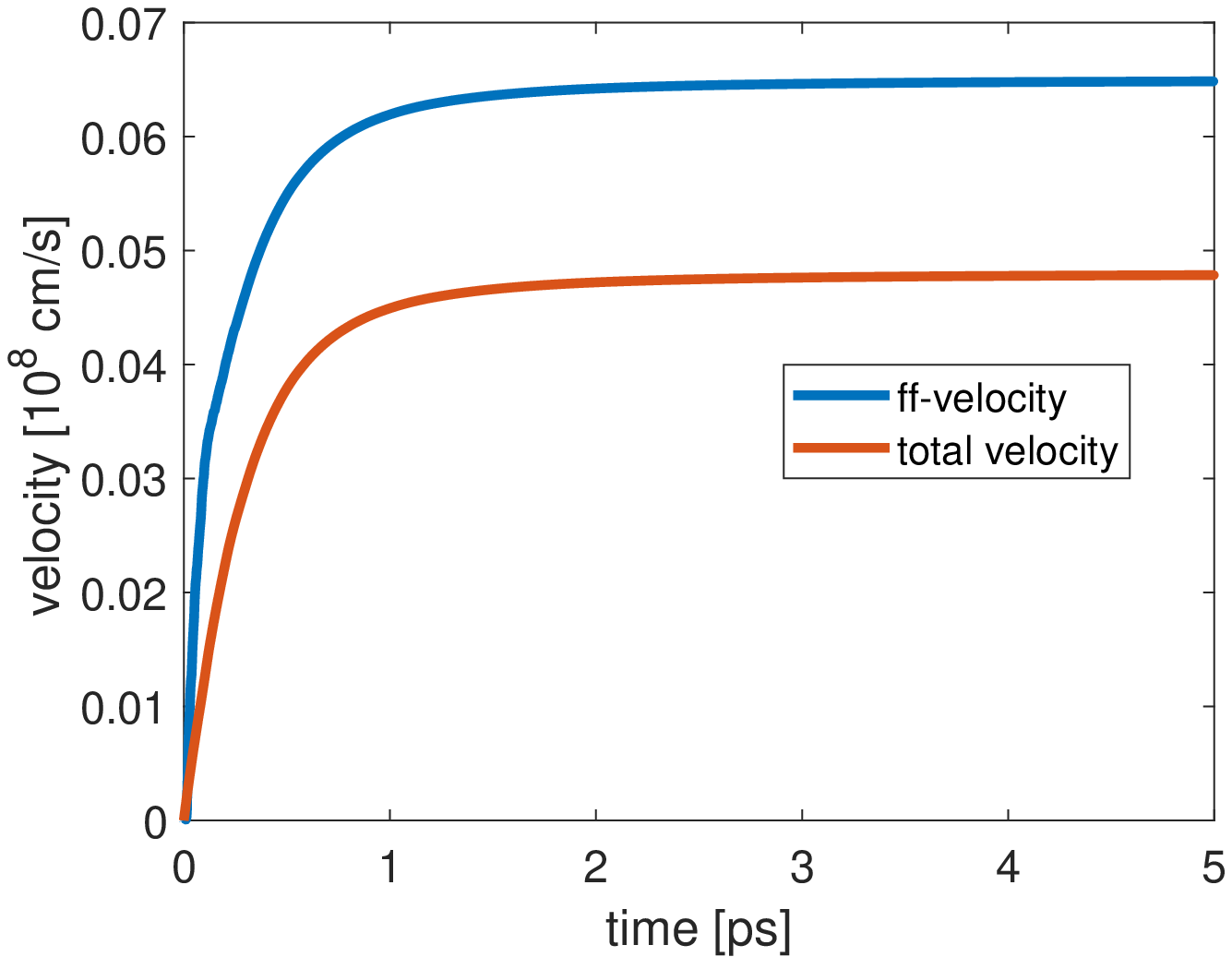}}
	\fbox{b)			\includegraphics[width=0.41\columnwidth]{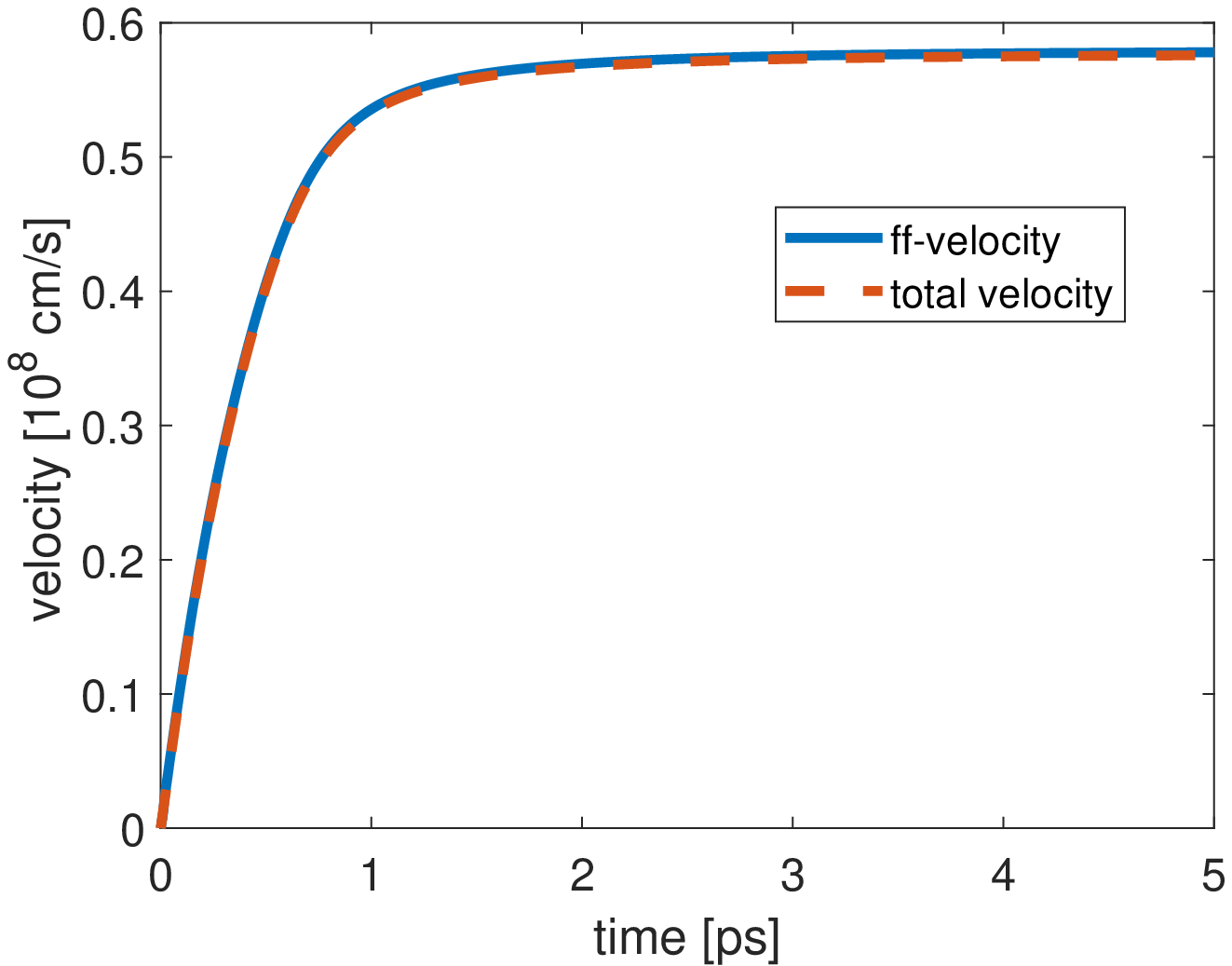}}\\
	\caption{Total mean velocity and velocity due to the free flight, when all the scatterings are neglected, for the FFMC with the FD, a), and the MB, b), initial condition, respectively.	\label{ff_vel}}
\end{figure}

The previous results show that with FFMC there are really few scattering events; their contribution produces a negative final velocity which cannot be balanced by the free flights because these are almost completely inhibited and are not statistically relevant. This interpretation is confirmed by Figs. \ref{LO_K_vel} and \ref{ff_vel}.  In Fig. \ref{LO_K_vel}, the mean velocity is shown when only the scattering events with the longitudinal optical ($LO$) and the $K$ phonons, respectively, are considered. It is negative in both the cases and has a higher absolute value when only the $K$-phonons are included. 
In Fig. \ref{ff_vel} the mean velocity due to only the free flights, without any scattering, is shown. In Fig.\ref{ff_vel} a) the initial Fermi-Dirac distribution is used and the mean value is calculated by considering all the particles of the sample (rd line) and only those particles which experience a free flight event at least once (blue line), respectively. 
The two lines are very different, which means that a large number of particles does not experience any free flight during the simulation. The use of the initial MB in place of the FD distribution worsens the results because the mean negative velocity is larger in absolute value with respect to the FD. This is due to the fact that in the early times of the simulation there are more available states and each particle has a free flight event at least once, so that the previous two samples are almost the same and the corresponding mean velocities are almost equal, as it is evident in Fig. \ref{ff_vel} (b). 

\begin{figure}[h]
	\centering
	\includegraphics[width=0.6\columnwidth]{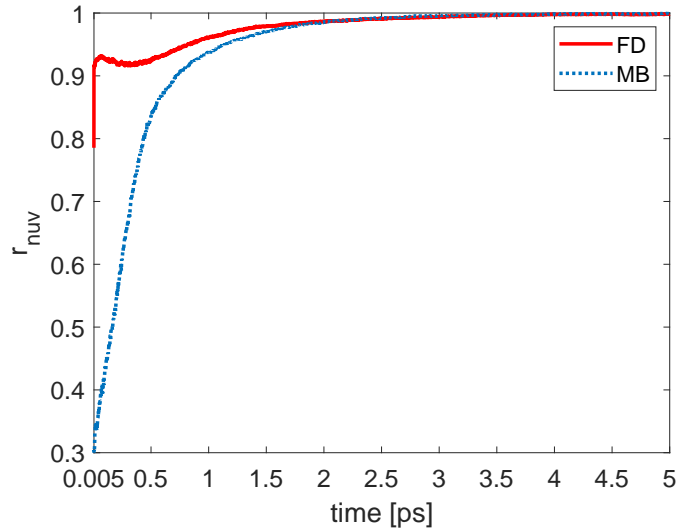}\\
	\caption{Percentage of particles that do not vary their velocity at each time step for the FFMC, with the FD, a), and the MB, b), initial condition, respectively.    \label{nvn}}
\end{figure}

When a particle experiences a change of state due to either free flight or scattering with the crystal lattice phonons, it keeps moving in a free motion, without changing its velocity during the rest of the simulation; when this velocity is negative, it will remain negative until the simulation has finished. This fact is supported by the count of the percentage of particles that do not change their velocity between two consecutive time steps, as reported in Fig. \ref{nvn}; after about $1.5$ ps, almost all the particles perform only a free motion.

\begin{figure}[h]
	\centering
	\fbox{a)			\includegraphics[width=0.41\columnwidth]{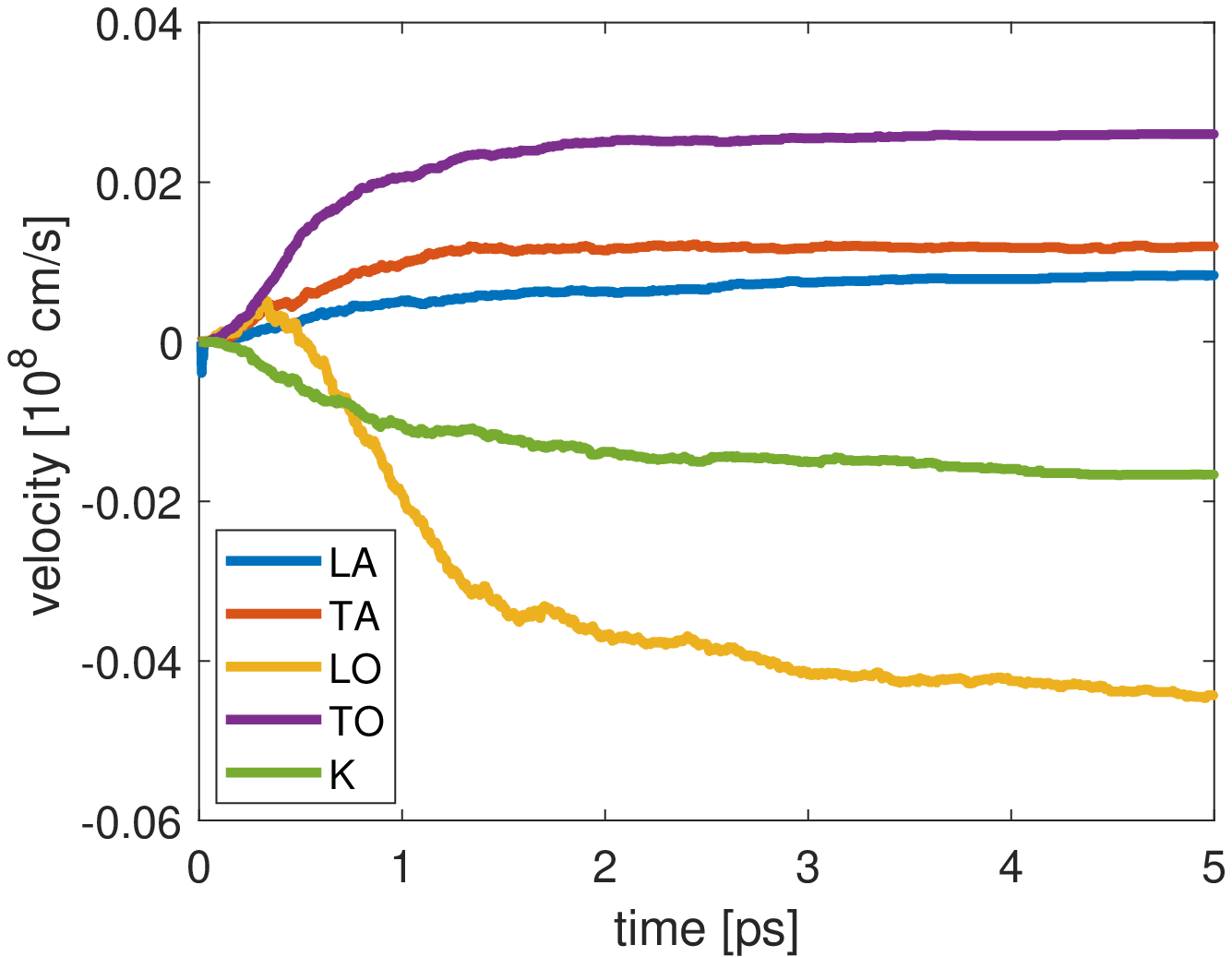}}
	\fbox{b)			\includegraphics[width=0.41\columnwidth]{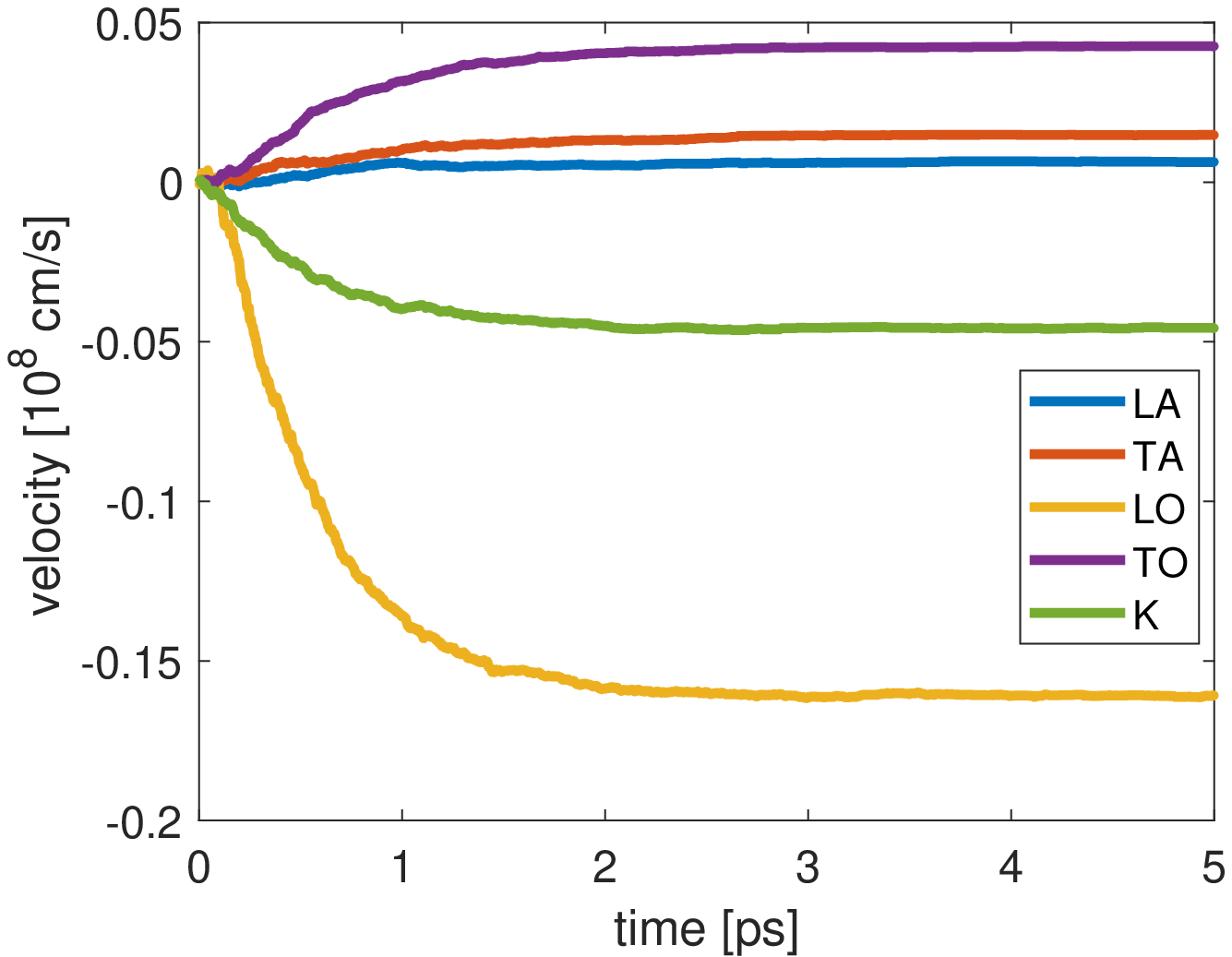}}\\
	\caption{Contribution of each scattering type on the mean velocity for the FFMC with the FD, a), and the MB, b), initial condition, respectively.	\label{sc_vel}}
\end{figure}

In Fig. \ref{sc_vel}, the contribution of each type of scattering to the mean velocity, obtained by averaging only over the particle population which experienced the corresponding collision events, is shown. The prevalence of the backward scatterings with $LO$ and $K$ phonons is confirmed; the $LA$, $TA$ and the transversal optical ($TO$) phonons give a positive contribution, but their number is too low to balance the negative values due to the $LO$ and $K$ collisions. 

\section{Conclusions}
\label{conclusions}
The inclusion of the Pauli principle in the Monte Carlo simulations by introducing the rejection technique also at the end of each free flight, and not only after a scattering event, is analyzed, by considering a homogeneous suspended monolayer of graphene as test case. The results are compared with those obtained when the Pauli principle is applied only at the end of the collision events, both in the standard Ensemble Monte Carlo and in the new Ensemble Monte Carlo scheme, the latter being able to take into account the Pauli principle correctly. The treatment of the free flight in a ``quantum'' perspective is not in accordance with the Liouvillian nature of the streaming part of the Boltzmann equation. By employing this approach, the resulting numerical dynamics does not seem compatible with the solution of the Boltzmann equations and leads to unphysical results. This is due to the fact that the most part of the free flights are rejected, the statistics is very poor and becomes no longer significant and the distribution function changes only very little with respect to the initial one. The mean value of the energy is almost constant and the mean velocity is negative, due to the effect of the backwards scatterings, mainly with the longitudinal optical and the $K$ phonons which are not balanced by the free flights or by the forward scatterings. The same results are obtained also when the initial Fermi-Dirac distribution is substituted by a Maxwell-Boltzmann at high temperature; in this case a greater initial possibility of movement is present but the charge distribution tends to acquire an irregular Fermi-Dirac-like shape in the course of its evolution. The average values of velocity are even worse, negative and larger in absolute value. The average energy starts from a high value and rapidly decreases to a stationary one, in accordance with the evolution of the distribution function. The statistical sample is more populated at early times but it quickly becomes too poor. For both initial distributions almost all the particles that experience a change of state will not change their velocity during the whole simulation so maintaining their state. The FFMC procedure does not reproduce a solution of the Boltzmann equation when degeneracy effects are relevant.\\

\begin{acknowledgments}
M.C. acknowledges the research fellowship of Department of Industrial Engineering and Mathematical Sciences of Polytechnic University of Marche, ``DIPARTIMENTI DI ECCELLENZA -DIISM - Responsabile Prof. Michele Germani dal $01.11.2020$ – (040004-MIUR-DIP-ECCELL-2018) – CUP I31G18000030005'', and ``Progetto Giovani 2019'' by GNFM (INdAM). M.C. and V.R. acknwoledge the support by GNFM (INdAM).
\end{acknowledgments}


\begin{thebibliography}{99}
	
	\bibitem{LP}
	P.A. Lebwohl and P.J. Price, ``Hybrid Method for Hot Electron Calculations'', Solid State Communications Vol. 9, pp. 1221—1224, 1971.
	
	\bibitem{BJ}
	S. Bosi and C. Jacoboni, ``Monte Carlo high-field transport in degenerate GaAs'', J. Phys. C: Solid State Phys. {\textbf{9}} 315, 1976.
	
	\bibitem{JacReg}
	C. Jacoboni and L. Reggiani, ``The Monte Carlo method for the solution of charge transport in semiconductors with applications to covalent material'', Rev. Mod. Phys., vol. 55, pp. 645-705, July 1983.
	
	
	\bibitem{LuFe}
	P. Lugli and D. K. Ferry,``Degeneracy in the ensemble Monte Carlo method for high-field transport in semiconductors'', IEEE Transactions on Electron Devices, vol. 32, no. 11, pp. 2431-2437, Nov. 1985, doi: 10.1109/T-ED.1985.22291.
	
	%
	%
	
	\bibitem{BoTho} 
	P. Borowik, J.~L. Thobel, Improve Monte Carlo method for the study of electron transport in degenerate semiconductors, J. Appl. Phys. 84 (1998) 3706.
	
	\bibitem{BoAda} 
	P. Borowik, L. Adamowciz, Improved algorithm for Monte Carlo studies of electron transport in degenerate semiconductors, Physica B 365  (2005) 235--239..
	
	\bibitem{FisLax} 
	M.~V. Fischetti, S.~E. Laux, Monte Carlo analysis of electron transport in small semiconductor devices  including band-structure and space-charge effects, Phys. Rev. B 38 (1988) 9721
	
	\bibitem{ZeBuEsSh}
	M. Zebarjadi, C. Bulutay, K. Esfarjani, A. Shakouri, Monte Carlo simulation of electron transport in degenerate and inhomogeneous semiconductors, Appl. Phys. Letters 90 (2007) 092111. 
	
	\bibitem{Tady}
	P. Tadyszak, F. Danneville, A. Cappy, L. Reggiani, L. Varani, and L. Rota, ``Monte Carlo calculations of hot-carrier noise under degenerate conditions'', Applied Physics Letters 69, 1450 (1996); doi: 10.1063/1.117611.
	
	\bibitem{1}
	Ferry D.K., Goodnick S.M. (2001) Ensemble Monte Carlo Simulations of Ultrafast Phenomena in Semiconductors. In: Tsen KT. (eds) Ultrafast Phenomena in Semiconductors. Springer, New York, NY. 
	
	
	\bibitem{3}
	A. Islam, K. Kalna, Monte Carlo simulations of mobility in doped GaAs using self-consistent Fermi–Dirac statistics,2011 Semicond. Sci. Technol. 26 055007.
	
	\bibitem{RMC} 
	Romano, V., Majorana, A., and Coco, M., 2015, ``DSMC Method Consistent With the Pauli Exclusion Principle and Comparison With Deterministic
	Solutions for Charge Transport in Graphene'' J. Comput. Phys., 302, pp. 267–284.
	
	\bibitem{CMR_Ric}
	M. Coco, A. Majorana, V. Romano. ``Cross validation of discontinuous Galerkin method and Monte Carlo simulations of charge transport in graphene on
	substrate''. Ricerche Mat. (2017) 66:201-220.
	
	\bibitem{CMNR_AAPP} 
	M. Coco, A. Majorana, G. Nastasi, V. Romano. ``High-field mobility in graphene on substrate with a proper inclusion of the Pauli exclusion principle''. Atti della
	Accademia Peloritana dei Pericolanti, Classe di Scienze Fisiche, Matematiche e
	Naturali, ISSN 1825-1242, Vol. 97, No. S1, A6 (2019), DOI:
	10.1478/AAPP.97S1A6.
	
	\bibitem{NR}
	G. Nastasi, V. Romano. Improved mobility models for charge transport in graphene. Commun. Appl. Ind. Math., vol 10, pp. 41-52, May 2019.
	
	\bibitem{NR2}
	G. Nastasi, V. Romano. A full coupled drift-diffusion-Poisson simulation of a GFET. Commun. Nonlinear Sci. Numer. Simulat., vol 87, Aug. 2020.
	
	\bibitem{CN}
	Coco M., Nastasi G. ``Simulation of bipolar charge transport in graphene on h-BN'', COMPEL - The international journal for computation and mathematics in electrical and electronic engineering, Vol. 39 No. 2, pp. 449-465, 2020. https://doi.org/10.1108/COMPEL-08-2019-0311
	
	\bibitem{MNR}
	A. Majorana, G. Nastasi, V. Romano. ``Simulation of Bipolar Charge Transport in Graphene by Using a Discontinuous Galerkin Method''. Comm. Comput. Phys., Vol.
	26, no 1, pp. 114-134, Feb. 2019.
	
	\bibitem{LiMo}
	P. Lichtenberger, O. Morandi, F. Schürrer, High-field transport and optical phonon scattering in graphene, Phys. Rev. B 84 (2011) 045406
	
	\bibitem{CaNe}
	Castro Neto, A. H., Guinea, F., Peres, N. M. R., Novoselov, K. S., Geim, A. K. (2009).
	The electronic properties of graphene.
	{\it Rev. Mod. Phys.} 81: 109--162.
	
	\bibitem{CR}
	M. Coco, V. Romano. ``Simulation of electron-phonon coupling and heating
	dynamics in suspended monolayer graphene including all the phonon branches''.
	J. Heat Transfer. 2018; 140(9):092404-092404-10. doi:10.1115/1.4040082.
	
	\bibitem{MR}
	G. Mascali, V. Romano. Charge transport in graphene including thermal effects. SIAM J. Appl. Math., vol. 77, no 2, pp. 593-613, Apr. 2017.
	
	\bibitem{CMR_16}
	M. Coco, G. Mascali, V. Romano. ``Monte Carlo analysis of thermal effects in
	monolayer graphene''. Journal of Computational and Theoretical Transport, Vol.
	45 (7) : 540-543 (2016).
	
	\bibitem{CR_19}
	M. Coco, V. Romano. ``Assessment of the constant phonon relaxation time
	approximation in electron-phonon coupling in graphene''. Journal of
	Computational and Theoretical Transport. 7 (1-3), 246–266, 2018. doi:
	10.1080/23324309.2018.1558253.
	
	\bibitem{JL}
	C. Jacoboni, P. Lugli, The Monte Carlo Method for Semiconductor Device Simulation, Springer-Verlag, Berlin, 1989.
	
	\bibitem{Sarma} 
	Das Sarma, S., Adam, S., Hwang, E. H., Rossi, E. (2011). Electronic transport in two-dimensional graphene. {\it Review of Modern Physics} 83: 407-407.
	
	\bibitem{Li2010}
	X. Li, E.~A. Barry, J.~M. Zavada, M. Buongiorno Nardelli, K.~W. Kim,
	Surface polar phonon dominated electron transport in graphene.
	Appl. Phys. Lett., {\bf 97},  232105 (2010)
	
	\bibitem{Bor} K.~M. Borysenko, J.~T. Mullen, E.~A. Barry, S. Paul, Y. ~G. Semenov, J.~M. 
	Zavada, M. Buongiorno Nardelli, K.~W. Kim,
	First-principles analysis of electron-phonon interactions in graphene.
	{Phys. Rev. B} {\bf 11},  121412(R) (2010)
	
	
	
	
	
\end{thebibliography}
\end{document}